\documentclass[a4paper,fleqn]{cas-dc}

\usepackage[authoryear]{natbib}

\usepackage{bm} 

\def\tsc#1{\csdef{#1}{\textsc{\lowercase{#1}}\xspace}}
\tsc{WGM}
\tsc{QE}
\tsc{EP}
\tsc{PMS}
\tsc{BEC}
\tsc{DE}

\begin{document}
\let\WriteBookmarks\relax
\def\floatpagepagefraction{1}
\def\textpagefraction{.001}

\shorttitle{ }

\shortauthors{K. Deng et~al.}

\title [mode = title]{Decoupled Structure for Improved Adaptability of End-to-End Models}                      



%
\author{Keqi Deng}
\ead{kd502@cam.ac.uk}
\author{Philip C. Woodland}
\cormark[1]
\tnotetext[1]{Corresponding author.}


\ead{pcw@eng.cam.ac.uk}



\affiliation{organization={Department of Engineering, University of Cambridge},
    addressline={Trumpington Street}, 
    city={Cambridge},
    postcode={CB2 1PZ}, 
    country={UK}}










\begin{abstract}
Although end-to-end (E2E) trainable automatic speech recognition (ASR) has shown great success by jointly learning acoustic and linguistic information, it still suffers from the effect of domain shifts, thus limiting potential applications. The E2E ASR model implicitly learns an internal language model (LM) which characterises the training distribution of the source domain, and the E2E trainable nature makes the internal LM difficult to adapt to the target domain with text-only data.
To solve this problem, this paper proposes decoupled structures for attention-based encoder-decoder (Decoupled-AED) and neural transducer (Decoupled-Transducer) models, which can achieve flexible domain adaptation in both offline and online scenarios while maintaining robust intra-domain performance.
To this end, the acoustic and linguistic parts of the E2E model decoder (or prediction network) are decoupled, making the linguistic component (i.e. internal LM) replaceable. When encountering a domain shift, the internal LM can be directly replaced during inference by a target-domain LM, without re-training or using domain-specific paired speech-text data. Experiments for E2E ASR models trained on the LibriSpeech-100h corpus showed that the proposed decoupled structure gave 15.1$\%$ and 17.2$\%$ relative word error rate reductions on the TED-LIUM 2 and AESRC2020 corpora while still maintaining performance on intra-domain data.
\end{abstract}




\maketitle

\section{Introduction}
The hybrid deep neural network and hidden Markov model (DNN-HMM) framework \citep{Hinton2012DeepNN, 5740583} 
is a widely used
deep learning-based approach for automatic speech recognition (ASR).
The hybrid DNN-HMM  
contains several separately optimised modules \citep{LI202312} including the acoustic model, the pronunciation lexicon, the context dependency model \citep{Young1994TreeBasedST}, and the language model (LM), which each uses different training objective functions.
However, this makes it hard to achieve overall system optimality \citep{Wang2019AnOO}.
End-to-end (E2E) trainable ASR models integrate the modules used by hybrid DNN-HMM ASR methods into one \citep{6638947,9688009} model and directly transcribe input speech into output transcripts. E2E ASR models such as the attention-based encoder-decoder (AED) \citep{Chan2016ListenAA} and the neural transducer \citep{Graves2012SequenceTW}
jointly learn acoustic and linguistic information \citep{LI202312} and predict words
directly without a separate lexicon and context dependency model and hence simplify the decoding process. 

Due to the availability of large-scale labelled data, the word error rate (WER) result of
E2E ASR surpasses conventional hybrid DNN-HMM methods on most public corpora \citep{9688009}. However, E2E models still suffer from domain shift issues between training and testing \citep{Non-Parametric2022, tsunoo22_interspeech}, and it's not always feasible to collect
a large quantity of target-domain speech-text paired data and hence it may be limited in quantity \citep{9746480}. In contrast, a target-domain text-only corpus is usually easier to obtain, and it is more efficient to bias E2E ASR models toward the target domain
using only such data \citep{keqi2023rilm, tsunoo22_interspeech}.

Previous work addressing the domain shift problem using text-only data mainly falls into three categories: external LM fusion; internal LM estimation; and text-to-speech (TTS) based methods. For external LM fusion, shallow fusion \citep{chorowski2015attention}, which linearly interpolates E2E ASR model scores (i.e. log probabilities)
with those from an external LM, is straightforward and widely deployed \citep{8462682}. Several structural fusion methods such as deep fusion \citep{gulcehre2015using}, cold fusion \citep{sriram18_interspeech}, and component fusion \citep{Shan2019ComponentFL} have also been proposed, but they require additional training and have not replaced shallow fusion as the dominant method for LM integration \citep{9003790, 9383515}. 

E2E ASR models implicitly learn an internal LM \citep{9383515} which characterises the training distribution of the source domain. There have been several studies concerning internal LM estimation \citep{9003790, Variani2020HybridAT, zeineldeen21_interspeech, 9415039, 9383515, 9746948}.
A density ratio method \citep{9003790} was introduced as an extension of shallow fusion, which
estimates the score from a separate source-domain LM that is to be subtracted from the target-domain LM score. 
HAT \citep{Variani2020HybridAT} was proposed as an efficient way to estimate the internal LM by removing the effect of the encoder from the transducer network.
However, these methods complicate the decoding process and an accurate estimate of the internal LM is not always feasible due to domain mismatch \citep{tsunoo22_interspeech}. In addition, recent work \citep{Chen2021FactorizedNT, Meng2022ModularHA, meng22_interspeech} such as the factorised neural transducer \citep{Chen2021FactorizedNT}
focuses on fine-tuning the internal LM on target-domain text, 
which can degrade intra-domain performance \citep{Chen2021FactorizedNT} or rely on Kullback-Leibler divergence regularisation that avoids this issue but limits how well the internal LM can learn the target domain \citep{Meng2022ModularHA, meng22_interspeech}.

With the development of high-quality neural
TTS, a new trend is to adapt E2E ASR models with the synthesised speech generated from the target-domain text data \citep{Zheng2020UsingSA, Peyser2019ImprovingPO}, but training a high-quality TTS model is expensive \citep{SIP-2021-0050} and the TTS speech still differs from natural human speech thus under the risk of performance degradation on human speech \citep{Li2019SemisupervisedTF}.

Domain adaptation is not such a severe issue for the conventional hybrid DNN-HMM method \citep{SIP-2021-0050} since an explicit independent LM is used.
However, compared to the E2E model, the DNN-HMM method optimises individual components separately rather than the joint optimisation in the E2E which can lead to a less-well optimised overall system and may also suffer from error propagation issues \citep{SIP-2021-0050}. In order to retain the advantages of flexible adaptation from conventional hybrid DNN-HMM methods in an E2E model, this paper proposes an E2E model structure that decouples the acoustic and linguistic parts of the E2E model decoder in the AED or the prediction network in the neural transducer.  In order to maintain the advantage of E2E models of optimising the entire model with a task-consistent objective, this decoupled structure still follows the E2E training approach, but the acoustic and linguistic information are jointly combined in a more modular way (i.e. addition of logits). 
Therefore, the proposed decoupled structure combines the advantages of the conventional DNN-HMM method and E2E trainable models.
Overall, the main contributions of our work are summarised in three key aspects.

First, the proposed decoupled structure is incorporated into the
attention-based encoder-decoder (AED) approach and denoted the Decoupled-AED. In the Decoupled-AED, the cross-attention modules of the Transformer \citep{Vaswani2017} decoder are decoupled from the self-attention modules, since it is the self-attention that enables the decoder to model the context between output tokens and therefore is responsible for the operation of
the internal LM.
In the Decoupled-AED, the linguistic component (i.e. the internal LM) can be replaced by a target-domain LM during inference if 
there is a domain shift. The target-domain LM only requires text data
 and the E2E ASR model doesn't require re-training.

The proposed decoupled structure is then extended to the neural transducer structure which is called the Decoupled-Transducer in this paper. 
The Decoupled-Transducer is evaluated in both offline and online scenarios, in which a chunk-based online fine-tuning strategy is implemented for self-supervised pre-trained models.

Finally, extensive experiments across different model structures, datasets and tasks have been carried out to evaluate the proposed decoupled structure. 
A further extension of the decoupled structure
has also been explored for the E2E speech translation (ST) task. 

Experiments with ASR models that were trained on the LibriSpeech-100h \citep{7178964} corpus show that the proposed decoupled structure greatly boosts cross-domain ASR accuracy 
while maintaining competitive intra-domain results.

The rest of this paper is organised as follows. 
Section~\ref{bac} introduces background work.
Section~\ref{method} details the proposed decoupled structure based on AED and transducer models.
The experimental setups and results are shown in Sections~\ref{setup} and~\ref{result} respectively. Finally, the paper concludes in Section~\ref{conclusion}.
\vspace{-0.5cm}
\section{Background}
\label{bac}
This section reviews the AED model, CTC/attention joint recognition, the neural transducer, and the pre-trained Transformer. This background information is referred to throughout this paper. 

\subsection{Attention-based Encoder-decoder Models}
There are several variants (e.g. \cite{Chan2016ListenAA, 8682586}) of the AED model used for ASR, and this paper focuses on the Transformer-based AED structure \citep{Vaswani2017} which recently has been widely studied. 
The main difference between the Transformer and
other AED models is that the Transformer is solely based on the attention mechanism, without using a recurrent neural network (RNN) \citep{Vaswani2017}. The attention mechanism maps a query vector and a set of key-value vector pairs to an output vector via scaled dot-product attention, which is used as the basic attention function:
\begin{equation}
    {\rm Attention}\left( {\textbf{Q},\textbf{K},\textbf{V}} \right) = {\rm softmax}\left( {\frac{\textbf{Q}\textbf{K}^{T}}{\sqrt{d_{k}}}} \right)\textbf{V}
\end{equation}
where the matrices $\mathbf{Q}$, $\mathbf{K}$, and $\mathbf{V}$ refers to the queries, keys, and values respectively and $d_{k}$ is the dimension of the keys.

The AED model directly maps a $T$-length sequence of
input speech $\bm{x}$ into a $N$-length target text sequence $\bm{y}$ using an encoder-decoder structure. 
The posterior distribution computed by the AED model follows the chain rule of conditional probability:
\begin{equation}
    p(\bm{y}|\bm{x}) = \prod_{n}p(y_n|\bm{x}, \bm{y}_{1:n-1})
\end{equation}
The encoder converts the input speech into an acoustic representation $\mathbf{H}^{\rm enc}=(\bm{h}_1^{\rm enc}, \cdots, \bm{h}_T^{\rm enc})$ and feeds it to the decoder, which jointly learns acoustic and linguistic information and predicts the next element of the sequence as:
\begin{equation}
    p({y}_n|\bm{x}, \bm{y}_{1:n-1}) = {\rm TransformerDecoder}(\bm{y}_{1:n-1}, \mathbf{H}^{\rm enc}) \label{dec-p}
\end{equation}

The Transformer encoder and decoder both contain several identical layers.
To be more specific, the Transformer encoder layer is based on a stack of feed-forward modules on top of a self-attention module which performs
multi-head attention over the encoder input. Compared to the encoder, the Transformer decoder layer inserts an additional cross-attention module between the self-attention and feed-forward modules to perform multi-head attention over the encoder output and the previous layer’s output. The decoder self-attention module performs multi-head attention over the previous tokens or over the output of the previous decoder layer. To prevent the decoder from seeing future information in the context and to preserve the auto-regressive property, future tokens are masked \citep{Vaswani2017} for the self-attention module which makes the decoder unidirectional.
\subsection{Connectionist Temporal Classification (CTC)}
CTC \citep{graves2006connectionist} was the first E2E technology widely used in ASR. CTC 
considers all possible alignments between the input speech sequence and output text token sequences \citep{SIP-2021-0050}. To align these sequences at the frame level, a blank label is inserted between tokens while allowing repetition of the same tokens \citep{10.5555/3044805.3045089}.
Denoting the input speech frames as $\bm{x}$, target text as $\bm{y}$, $A^{-1}(\bm{y})$ is all possible CTC alignments mapped from $\bm{y}$. The CTC loss function is defined as the negative log probabilities of target text given the input speech:
\begin{equation}
    L_{\rm ctc}=-{\rm ln}\sum_{\bm{q} \in A^{-1}(\bm{y})}p(\bm{q}|\bm{x})
\end{equation}
where $\bm{q}$ is a possible CTC path. Under a conditional independence assumption between the output tokens, $p(\bm{q}|\bm{x})$ can be expressed as:
\begin{equation}
    p(\bm{q}|\bm{x}) = \prod_{t=1}^{T}p(q_t|\bm{x})
\end{equation}
where T is the length of input speech and $p(q_t|\bm{x})$ is the predicted probability at the $t$-th frame that can be computed by applying the softmax function to the logits output by the encoder, which is similar to that of AED or transducer.

CTC trains the encoder with the blank label which contains no linguistic information using the forward-backward algorithm \citep{graves2006connectionist}. 
It can be shown, e.g. \cite{9003837, LI202312}, that CTC is actually
equivalent to a special instantiation of the two-state HMM structure when prior and transition probabilities are constant for any state.

\subsection{Neural Transducer Models}
The neural transducer \citep{Graves2012SequenceTW} provides a natural approach for online ASR \citep{SIP-2021-0050} and
contains an encoder network, a prediction network, and a joint network.
The encoder extracts an acoustic representation $\bm{h}_t^{\rm enc}$ from input speech. The prediction network generates a linguistic representation $\bm{h}_n^{\rm pre}$ from the previous non-blank output tokens $y_{1:n-1}$, which captures causal dependencies in the output 
\citep{Higuchi2022BERTMC}. The joint network combines $\bm{h}_t^{\rm enc}$ and $\bm{h}_n^{\rm pre}$ using fully-connected (FC) networks:
\begin{equation}
    \bm{l}_{t,n} = {\rm FC}(\Psi({\rm FC}(\bm{h}_t^{\rm enc})+{\rm FC}(\bm{h}_n^{\rm pre})))
\end{equation}
where $\Psi$ is a non-linear activation function and the predicted probability of token $k$ in the neural transducer can be computed by applying a softmax function to the logits $\bm{l}_{t,n}$:
\begin{equation}
    p(\hat{y}_{t+n}=k|\bm{x}_{1:t}, {y}_{1:n-1})={\rm softmax}(\bm{l}_{t,n})
\end{equation}
where $\hat{y}_{t+n}$ can be a blank token or a non-blank vocabulary token.
The neural transducer loss function $L_{\rm nt}$ is defined as the negative log likelihood of the token sequence:
\begin{eqnarray}
&p(\bm{a}|\bm{x})\approx\prod_{u=1}^{T+N}p(a_u|A(a_{1:u-1}),\bm{x})\\
  &L_{\rm nt}=-{\rm ln}\sum_{\bm{a} \in A^{-1}(\bm{y})}p(\bm{a}|\bm{x}) \label{nt_loss}
\end{eqnarray}
where $A$ is a function that maps all alignment paths $\bm{a}$ to the target text token sequence $\bm{y}$ of length $N$ by removing the blank token. The alignment paths are obtained using the forward-backward
algorithm.

The encoder network in a neural transducer uses a long short-term memory (LSTM) \citep{Hochreiter1997}, Transformer, or Conformer \citep{gulati20_interspeech} network. To enable streaming recognition, 
an RNN encoder needs to be unidirectional, and strategies like the chunk-based or the lookahead-based method \citep{li20_interspeech} need to be employed for a streaming Transformer encoder.
The prediction network normally contains an RNN \citep{Graves2012SequenceTW}, unidirectional Transformer \citep{Zhang2020TransformerTA} or even only an embedding layer which is called a stateless prediction network \citep{9054419}. 
The neural transducer has no independence assumptions between output symbols and can 
handle streaming speech data, making it the most popular E2E
model used in industry applications \citep{SIP-2021-0050}.

\subsection{CTC/attention Joint Training and Recognition}
Based on the standard AED, the CTC/attention joint model \citep{8068205} utilises CTC \citep{graves2006connectionist} to improve the model training and refine the beam search during ASR decoding. The CTC branch shares the encoder with an additional linear classifier, and the overall model is optimised via multitask learning:
\begin{equation}
    L_{\rm mtl}=\gamma L_{\rm ctc}+(1-\gamma)L_{\rm attention}
\end{equation}
During beam search, CTC/attention joint recognition considers both the attention-based decoder prediction and the CTC prefix score of the hypotheses. Suppose there is a $n$-length hypothesis generated by the decoder $\bm{y}=({y}_1, \cdots, {y}_n)$ and the score assigned by the decoder is:
\begin{equation}
    S_{\rm attention}=\sum_{i=1}^n {\rm log}p_{\rm att}(y_i|\bm{x}, \bm{y}_{1:i-1})
\end{equation}
where $p_{\rm att}(y_i|\bm{x}, \bm{y}_{1:i-1})$ is computed following
Eq.~\ref{dec-p}. The CTC prefix score is computed as:
\begin{equation}
    S_{\rm ctc}={\rm log}\sum_{j=i}^T p_{\rm ctc}(\bm{y}|\bm{h}_{1:j}^{enc})
\end{equation}
During CTC/attention joint decoding, beam search with a hyper-parameter $\mu$ is used to prune partial hypotheses in accordance with 
the scoring function:
\begin{equation}
    S=\mu S_{\rm ctc}+(1-\mu)S_{\rm attention} \label{joint-aed}
\end{equation}

\subsection{Pre-trained Transformer}
\label{preformer}
In the standard Transformer \citep{Vaswani2017} AED architecture, the Transformer decoder contains a stack of $N$ identical layers. Each layer consists of three modules: a self-attention module, a cross-attention module, and a feed-forward module. The cross-attention module makes the decoder dependent on the acoustic encoder output and thus cannot be separately pre-trained \citep{9688009}. 
The pre-trained Transformer (Preformer) \citep{9688009} modifies the Transformer decoder structure by removing the cross-attention modules from each of the $N$ layers and stacking the $N$ cross-attention modules after them. Denote the layer that consists of a self-attention module followed by a feed-forward module as a self-layer and the layer that contains only the cross-attention module as a cross-layer. The Preformer decoder is built by stacking $N$ cross-layers on top of $N$ self-layers. The $N$ self-layers of the decoder can then be separately pre-trained on text data or initialised by a pre-trained Transformer LM. The Preformer is an inspiration for this work and is compared in the experiment section.
\vspace{-0.1cm}
\section{Proposed Decoupled Structure}
\label{method}
This paper proposes a decoupled structure for E2E models to achieve flexible domain adaptation while maintaining good intra-domain performance. Applying the proposed decoupled structure to offline and online mainstream models, this method can be divided into the Decoupled-AED model and the Decoupled-Transducer model.
In this paper, the proposed decoupled structure uses an auxiliary CTC branch \citep{8068205, 9688251} to achieve more competitive performance. In this section,
 the challenge of domain adaptation in E2E ASR is first introduced.
Then, the proposed Decoupled-AED and Decoupled-Transducer are described.
\begin{figure*}[t]
    \centering
    \includegraphics[width=0.95\linewidth]{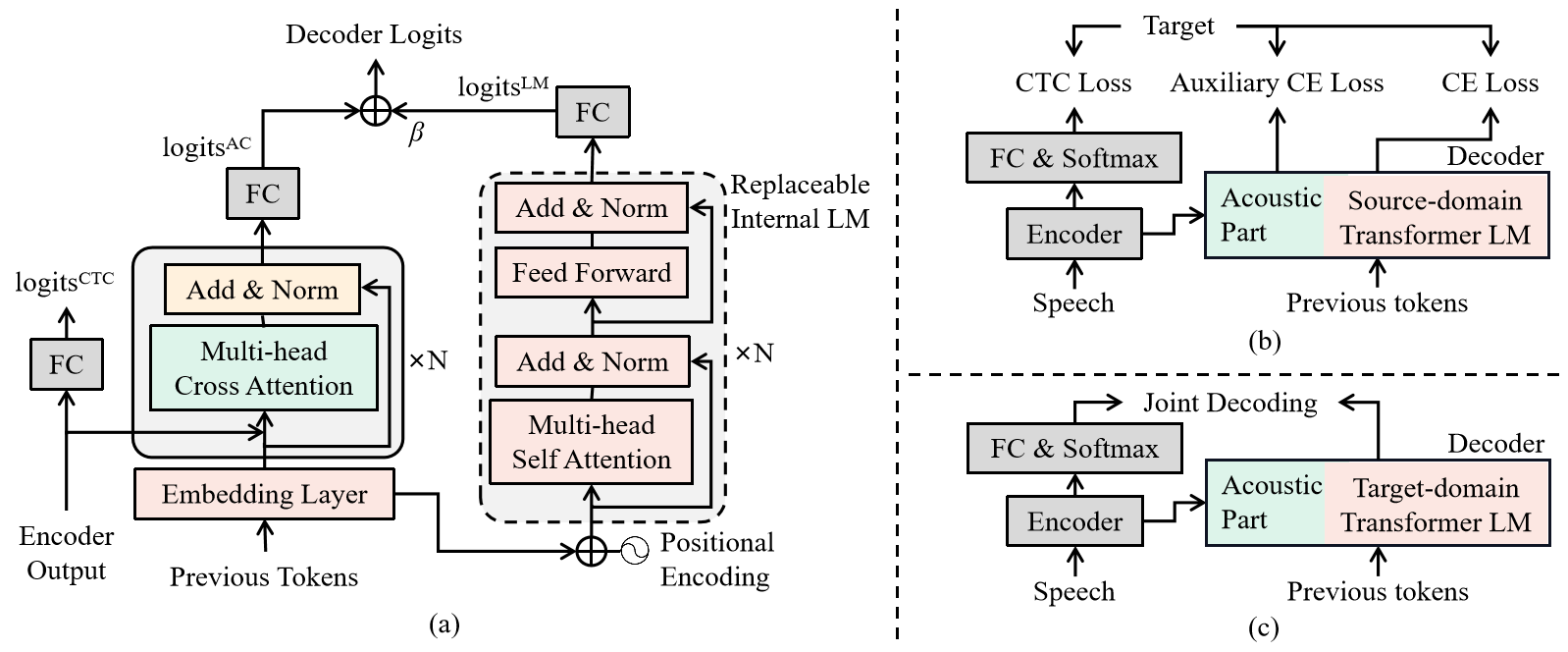}
    \caption{Illustration of the Decoupled-AED structure. (a) decoder in the Decoupled-AED; (b) the training process; (c) the decoding process. CE loss denotes the cross-entropy loss,
    FC represents a fully-connected layer, and \textcircled{+} denotes addition operations. The embedding layer is shared with the replaceable internal LM, which is Transformer LM in this paper. Therefore, the whole model only has one more FC layer than the standard AED model.}
    \label{fig:AED}
    \vspace{-0.3cm}
\end{figure*}
\subsection{Domain Adaptation in E2E Model}
Compared to conventional hybrid DNN-HMM models that separately optimise individual components,
E2E ASR models use a task-consistent objective function to optimise the whole network and achieve improved performance \citep{SIP-2021-0050}.
However, E2E ASR models jointly learn acoustic and linguistic information and the standard structure doesn't have an explicit separate LM
as used in the hybrid DNN-HMM approach. For example, in the Transformer decoder, although the self-attention modules can model the dependency between tokens,
its output will first be processed by the cross-attention module to be combined with the acoustic encoder output before passing it to the next self-attention module. This structure means that no part of the model can be explicitly regarded as representing the LM, which further leads to challenges in domain adaptation using text-only data. Another example is the prediction network in the neural transducer.
The prediction network needs to coordinate with the acoustic encoder to generate both blank and non-blank tokens \citep{Chen2021FactorizedNT}. However, the blank token is related to the acoustic input,
so the prediction network still cannot be considered as an explicit LM \citep{9054419, Chen2021FactorizedNT}.

Current E2E ASR models lack an explicit LM as part of the standard structure and thus domain adaptation using text-only data is challenging. However, in this paper, it is argued that E2E training and an explicit LM structure are not contradictory, and having an explicit part of the model that represents the LM does not need to lead to performance (especially intra-domain performance) degradation but enables flexible domain adaptation.

\begin{figure*}[t!]
    \centering
    \includegraphics[width=1\linewidth]{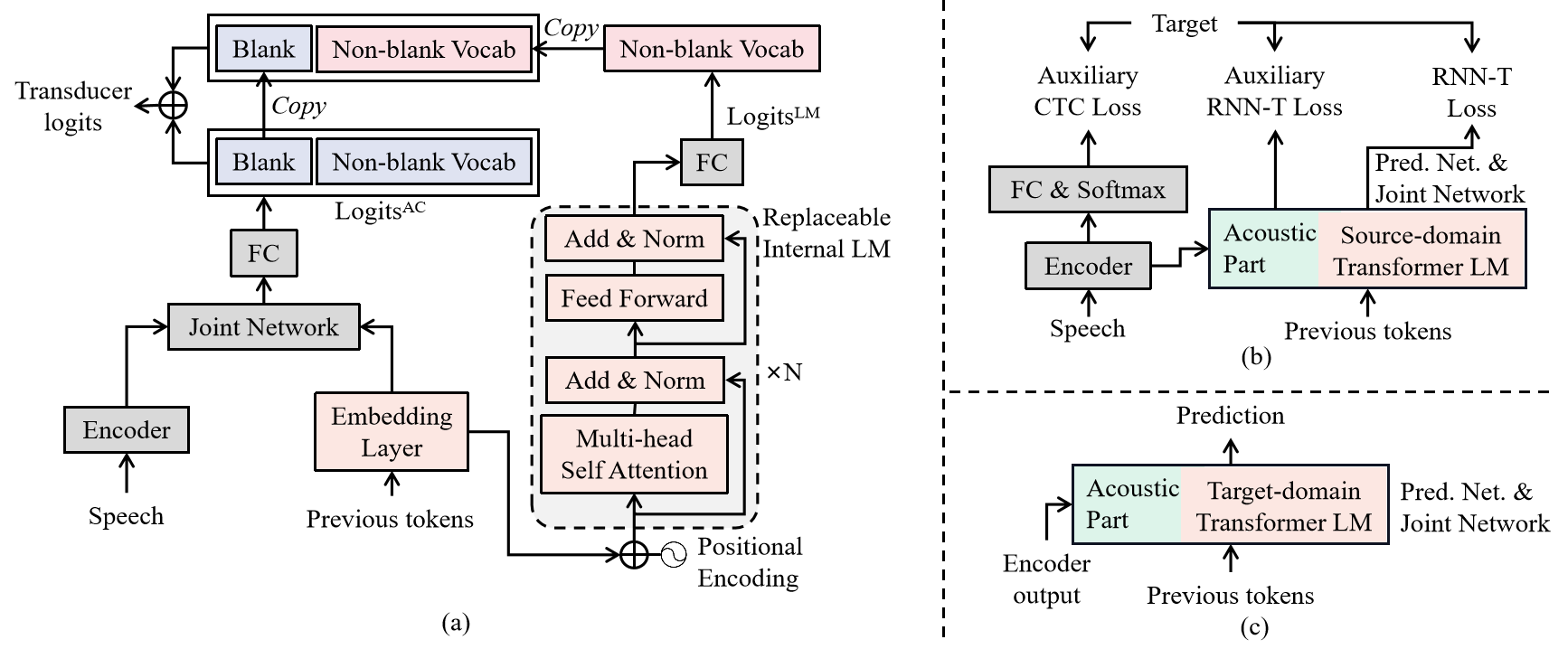}
    \caption{Illustration of the Decoupled-Transducer architecture. (a) the overall structure of the Decoupled-Transducer; (b) the training process; (d) the decoding process.
    The embedding layer is shared with the replaceable internal LM and the whole model only has one more FC layer than the normal neural transducer model.}
    \label{fig:NT}
    \vspace{-0.2cm}
\end{figure*}

\subsection{Decoupled-AED}
The decoupled structure proposed in this paper separates the decoder into acoustic and linguistic parts.
Since the self-attention module enables the Transformer decoder to model the context between output tokens and the cross-attention module allows the decoder to combine the acoustic encoder output,
the proposed structure separates the self-attention and cross-attention modules.

As shown on the left of Fig.~\ref{fig:AED}(a), the part of the model
inside the black solid line frame ($N$ cross-layers) is regarded as the acoustic part of the decoder, because it only has cross-attention modules to match the token embedding with the acoustic encoder output and cannot model the context between tokens. The replaceable internal LM (within the black dotted frame) on the right of Fig.~\ref{fig:AED}(a) is regarded as a linguistic part as it contains self-attention modules. This decoupled structure brings several benefits including (1) being more interpretable because acoustic and linguistic information are jointly learned through the addition of the logits; (2) the internal
LM is independent so that it can be trained or pre-trained on text-only data. Moreover, the parameters can be initialised directly using a pre-trained LM.
In this paper, this part of the proposed structure
employs a fixed LM that is pre-trained on source-domain text data during training to help the model converge. 
If a domain shift is encountered during decoding, the internal LM is directly replaced by a target-domain LM to tackle the domain mismatch.

The Decoupled-AED is trained as an E2E model, which retrains the advantage of optimising the entire model with an objective consistent with the task objective. 
To achieve improved results, the proposed Decoupled-AED employs the hybrid CTC/attention approach \citep{8068205}. Therefore, CTC supervision is used by the acoustic encoder by computing the CTC loss function ($L_{\rm ctc}$) between the CTC logits (i.e. ${\rm logits^{\rm CTC}}$ in Fig.~\ref{fig:AED}) and the target text token sequence. The logits output by the acoustic part of the decoder are denoted ${\rm logits}^{\rm AC}$ and those output by the linguistic part are denoted ${\rm logits}^{\rm LM}$. The final decoder logits are the weighted sum of ${\rm logits}^{\rm AC}$ and ${\rm logits}^{\rm LM}$:
\begin{equation}
    {\rm logits}^{\rm Dec} = {\rm logits}^{\rm AC}+\beta \cdot{\rm logits}^{\rm LM} \label{logits}
\end{equation}
where $\beta$ is a hyper-parameter. The decoder is trained using the cross-entropy (CE) loss ($L_{\rm ce}$) between ${\rm logits}^{\rm Dec}$ and the target text.

Furthermore, an auxiliary CE loss ($L_{\rm ce}^{\rm aux}$) is applied to the acoustic part (${\rm logits}^{\rm AC}$) to encourage the generation of a relatively accurate preliminary prediction based on acoustic information alone. Therefore, the overall objective of the Decoupled-AED is computed via the multi-task below:
\begin{equation}
    L_{\rm aed}=\gamma L_{\rm ctc}+(1-\gamma)(\eta L_{\rm ce}+(1-\eta)L_{\rm ce}^{\rm aux}) \label{multi-aed}
\end{equation}

Joint CTC/attention decoding \citep{8068205} is used during inference following Eq.~\ref{joint-aed}, which considers both the CTC prefix score of hypotheses and the decoupled decoder's beam search score that is computed as:
\begin{equation}
    S_{\rm attention}=-{\log}({\rm softmax}({\rm logits}^{\rm Dec}))
\end{equation}


\subsection{Decoupled-Transducer}
The prediction network of the neural transducer normally contains a recurrent neural network (RNN) \citep{Graves2012SequenceTW} or unidirectional Transformer \citep{Zhang2020TransformerTA}, and therefore explicitly captures the causal dependency between the output tokens and is a key difference to CTC \citep{Higuchi2022BERTMC}.
However, \cite{9054419} showed that the prediction network does not fully function as an LM, because it needs to predict both normal and blank tokens, which is inconsistent with the LM task \citep{SIP-2021-0050, Chen2021FactorizedNT}.
Therefore, this paper separates the non-blank token prediction from label prediction.

As shown in Fig.~\ref{fig:NT}(a), 
the embedding layer along with the encoder and joint networks 
is regarded as the acoustic part of the model, because it cannot explicitly model the dependency between output tokens through the Transformer or RNN and can only coordinate the acoustic encoder output with the current token.
Similar to the Decoupled-AED, the replaceable internal LM  (within the black dotted frame) on the right of Fig.~\ref{fig:NT}(a) is the linguistic part of the model. In the Decoupled-Transducer, the final logits (${\rm logits}^{\rm NT}$) are the sum of the logits based on acoustic information (${\rm logits}^{\rm AC}$) and the LM information (${\rm logits}^{\rm LM}$). Since the blank token is related to the alignment, it can be viewed as part of the acoustic model, and only the non-blank part of the ${\rm logits}^{\rm LM}$ is used. 
If the index of the blank label is $1$ and the vocab size is $V$, the process is as follows\footnote{This paper  multiplies ${\rm logits}^{\rm AC}_1$ by a factor of $2$ as ${\rm logits}^{\rm NT}_{1}$. However, excluding this factor is also valid, since the model will automatically learn to increase the value of ${\rm logits}^{\rm AC}_1$ during training to predict the blank token.}:
\begin{equation}
    \left\{
    \begin{array}{l}
        {\rm logits}^{\rm NT}_{1} = {\rm logits}^{\rm AC}_1+{\rm logits}^{\rm AC}_1 \\
        {\rm logits}^{\rm NT}_{2:V} = {\rm logits}^{\rm AC}_{2:V}+{\rm logits}^{\rm LM}_{2:V}
    \end{array}
    \right.
\end{equation}

Following the Decoupled-AED, the linguistic part of the model uses a fixed LM that is pre-trained on source-domain text data during training and is directly replaced by a target-domain LM when
decoding on an unseen domain. The Decoupled-Transducer is trained in an E2E fashion with the supervision of the neural transducer loss function ($L_{\rm nt}$) as described in Eq.~\ref{nt_loss}.
In addition, an auxiliary neural transducer loss ($L_{\rm nt}^{\rm aux}$) is computed based on the ${\rm logits}^{\rm AC}$ to generate a more accurate preliminary prediction based on the acoustic information alone. Furthermore, inspired by \cite{Zhao2022FastAA, 9688251},
an auxiliary CTC branch applied after the encoder is also used to improve the model convergence but discarded during inference. Therefore, the overall training objective of the Decoupled-Transducer is:
\begin{equation}
    L_{\rm nt}=\lambda L_{\rm ctc}+(1-\lambda)(\eta^{'}L_{\rm nt}+(1-\eta^{'})L_{\rm nt}^{\rm aux}) \label{ctc-nt}
\end{equation}
\section{Experimental Setup}
\label{setup}
\subsection{Datasets}
\label{dataa}
E2E ASR models were trained on the “train-clean-100” subset (LS100) of the LibriSpeech corpus \citep{7178964}, an audiobook corpus, and the standard dev/test sets from LibriSpeech (i.e. “dev-clean/-other” and “test-clean/-other”) were used for intra-domain evaluation. The text data for source-domain LM training were the training set transcripts and the LibriSpeech LM training text (40M sentences).
To show the effectiveness of the proposed decoupled structure for domain adaptation, 
two out-of-domain (OOD) datasets were employed in the experiments. The first OOD dataset was the TED-LIUM 2 \citep{rousseau-etal-2014-enhancing} dev/test sets, which is a spontaneous lecture style.
The text data for
target-domain LM training were the TED-LIUM 2 training set transcripts and corresponding LM training text (13M sentences).
The second OOD dataset was the AESRC2020 \citep{9413386} dev/test sets, which includes human-computer interaction (HCI) speech commands. The text data for target-domain LM training was its training set transcriptions. Details of the data are summarised in Table~\ref{corpus}.

The models and experimental evaluation were implemented based on the ESPnet \citep{Watanabe2018ESPnet} toolkit.
Raw speech data was used as input and 1000 modelling units as text output, including 997 BPE units \citep{gage1994} and 3 non-verbal symbols (i.e. blank, unknown-unit, and start/end-of-sentence).
\begin{table}[t] \footnotesize
\caption{Summary of datasets used for experiments}
\label{corpus}
\centering
\setlength{\tabcolsep}{1.4mm}
\renewcommand\arraystretch{1.4}
\begin{tabular}{ l|c|c }
\Xhline{2\arrayrulewidth}
 &\multicolumn{2}{c}{LibriSpeech-100h} \\
\hline
Style&\multicolumn{2}{c}{Audiobook Reading}\\
Training set &\multicolumn{2}{c}{train-clean-100}\\
\ \ \ -Total duration&\multicolumn{2}{c}{100 hours}\\
Intra-domain test sets&\multicolumn{1}{c}{dev-clean/-other} &test-clean/-other\\
\ \ \ -Total duration &\multicolumn{1}{c}{5.4/5.3 hours}&{5.4/5.1 hours}\\
\hline
\hline
&TED-LIUM 2& AESRC2020\\
\hline
Style&Spontaneous Lecture& HCI Command\\
Cross-domain test sets&test/dev &dev/test\\
\ \ \ -Total duration &2.6/1.3 hours& 14.5/21.0 hours\\
\Xhline{2\arrayrulewidth}
\end{tabular}
\end{table}
\subsection{Model Descriptions}
\subsubsection{AED Models}
The standard AED baseline model and the Decoupled-AED used the wav2vec2.0 encoder \citep{hsu21_interspeech} provided by Fairseq  (i.e. "w2v\_large\_lv\_fsh\_swbd\_cv") \citep{ott2019fairseq} and operate in the CTC/attention joint framework \citep{8068205}. The standard AED baseline (394.12M parameters) contained a 6-layer Transformer decoder with 1024 attention dimensions, 2048 feed-forward dimensions, and 8 heads. A Preformer \citep{9688009} baseline (394.12M parameters), as described in Sec.~\ref{preformer}, also used the wav2vec2.0 encoder and its decoder contained 6 self-layers and cross-layers, which had the same Transformer configuration (e.g. number of heads) as the standard AED baseline
The 6 self-layers of the Preformer decoder were initialised by a source-domain Transformer LM.
The replaceable internal LM of the Decoupled-AED (395.15M parameters) used a fixed source-domain Transformer LM and $N=6$ in Fig.~\ref{fig:AED}(a), and the Transformer configuration was the same as the standard AED baseline. The $\beta$ in Eq.~\ref{logits} was set to 0.5.
The $\gamma$ and $\mu$ for the CTC weight in Eq.~\ref{multi-aed} and Eq.~\ref{joint-aed} were set to $0.3$, while $\eta$ in Eq.~\ref{multi-aed} was set to 0.5.
The AED ASR models were trained for 35 epochs following the ESPnet2 recipe.
\subsubsection{Transducer Models}
Three neural Transformer Transducer (T-T) \citep{Zhang2020TransformerTA} baseline models were built and have different prediction networks. The neural T-T with an embedding layer as the prediction network is denoted as a stateless T-T (319.44M parameters), the T-T with a 6-layer 1024-dimensional LSTM prediction network \citep{Hochreiter1997} is denoted as LSTM T-T (369.82M parameters), and the T-T with a 6-layer unidirectional Transformer prediction network (1024 attention dimension, 2048 feed-forward dimensions, and 8 heads) is denoted as Transformer T-T (370.90M parameters). The replaceable internal LM used for the Decoupled-Transducer (371.92M parameters) was the same as that of the Decoupled-AED.
All of the baseline models and the Decoupled-Transducer used the wav2vec2.0 encoder which was the same as that used for AED models. For the online scenario, a chunk-based online fine-tuning strategy \citep{Cao2021ImprovingST} was implemented for wav2vec2.0 to yield a streaming wav2vec2.0 encoder. For this purpose,
a chunk-based mask \citep{li20_interspeech} was implemented for the encoder during training, with a 320~ms average latency.
The joint network dimension was 640. 
All models employed a
CTC branch to help training with a 0.3 weight, and
$\eta^{'}$ in Eq.~\ref{ctc-nt} was 0.5.
The Transducer models were trained for 25 epochs.
\subsubsection{External LM}
The source-domain 6-layer Transformer LM was built 
with a 1024 attention dimension, 2048 feed-forward dimension, and 8 heads.
It was trained for 25 epochs on the source-domain text data as described in Section~\ref{dataa} and fine-tuned on the target-domain text corpus for an extra 15 epochs as the target-domain LM. Shallow fusion \citep{chorowski2015attention} was implemented with a 0.2 LM weight if used for domain adaptation. A beam size of 10 was used during decoding. 
\section{Experimental Results}
\label{result}
Experiments were conducted to
compare the Decoupled-AED and Decoupled-Transducer with strong baseline models in both intra-domain and cross-domain scenarios.
\subsection{Experiments on Decoupled-AED}
\begin{table}[t] \footnotesize
    \caption{Intra-domain \%WER ($\downarrow$) results obtained on dev/test sets of LibriSpeech for AED ASR models trained on LibriSpeech 100-hour subset (LS100).}
    \label{tab:ls100}
  \centering
  \setlength{\tabcolsep}{2.3mm}
  \renewcommand\arraystretch{1.4}
  \begin{tabular}{l | c| c |c| c} 
    \Xhline{3\arrayrulewidth}
     \multirow{2}{*}{AED Models}&\multicolumn{2}{c|}{Test}&\multicolumn{2}{c}{Dev}\\
     \cline{2-5}
     &{clean}&{other}&{clean}&{other} \\
    \hline
    SpeechT5 \citep{Ao2021SpeechT5UE}&4.4 &10.4& 4.3&10.3\\
    Speech2C \citep{ao22_interspeech}&4.3&9.0&--&--\\
    UFO2 \citep{Fu2022UFO2AU}&5.0&11.8&--&--\\
    \hline
    Standard AED Baseline&6.4&8.1&5.6&8.3\\
    Preformer Baseline&4.5&7.3&4.1&7.4\\
    Decoupled-AED&\textbf{3.4}&\textbf{6.4}&\textbf{3.3}&\textbf{6.4}\\
    \Xhline{3\arrayrulewidth}
  \end{tabular}
\end{table}
Table~\ref{tab:ls100} lists the intra-domain word error rate (WER) results, which show
that our AED models achieved competitive performance with various recent results
on the LS100 benchmark. In addition, the Preformer \citep{9688009} baseline (as detailed in Sec.~\ref{preformer}) outperformed the standard AED baseline model because the Preformer baseline removed the cross-attention modules in the 6-layer Transformer decoder and stacked them at the end and initialises the parameters of the previous 6 layers by the pre-trained source-domain LM.
Furthermore, the Decoupled-AED achieved the best intra-domain performance, which might be due to the explicit use of the
pre-trained source-domain LM logits.
\begin{table}[t] \footnotesize
\caption{ASR \%WER ($\downarrow$) results in cross-domain adaptation scenario. TED-LIUM 2 data was abbreviated as Ted2 and AESRC2020 data was denoted as AESRC. SF denotes shallow fusion. Internal LM is the replaceable internal LM of the Decoupled-AED.}
  \label{tab:ls100cross}
  \centering
  \setlength{\tabcolsep}{2.0mm}
  \renewcommand\arraystretch{1.4}
  \begin{tabular}{l |c c|c c}
    \Xhline{3\arrayrulewidth}
     \multirow{2}{*}{AED Models}&\multicolumn{2}{c|}{LS100$\Rightarrow$Ted2}&\multicolumn{2}{c}{LS100$\Rightarrow$AESRC}\\
     &{~}{Test}&{Dev}{\;}&{~}{Dev}&{Test}{\;} \\
    \hline
    Normal AED Baseline&{~}10.8&11.4{\;}&{~}16.3&16.7{\;}\\
    \quad+Target-domain LM SF&{~}10.2&10.8{\;}&{~}14.6&14.8{\;}\\
    Preformer Baseline&{~}10.6&10.8{\;}&{~}16.2&16.9{\;}\\
    \quad+Target-domain LM SF&{~}9.7&10.1{\;}&{~}14.3&14.9{\;}\\
    \cline{1-1}
    Decoupled-AED&{~}9.8&10.5{\;}&{~}14.8&15.5{\;}\\
    +Replace Internal LM&{~}\textbf{9.0}&\textbf{9.6}{\;}&{~}\textbf{13.5}&\textbf{14.0}{\;}\\
    \quad++Target-domain LM SF&{~}8.6&9.1{\;}&{~}12.1&12.3{\;}\\
    \Xhline{3\arrayrulewidth}
  \end{tabular}
\end{table}

Experiments were then conducted to compare the cross-domain performance on the TED-LIUM 2 and AESRC2020 corpora. The cross-domain ASR WER results are shown in Table~\ref{tab:ls100cross}, in which the "LM SF" means using an external target-domain LM via shallow fusion. The results show that the cross-domain performance for the baseline models can be greatly improved with shallow fusion, although this came at the cost of extra computation and memory. However, the Decoupled-AED could  give
up to a further 9.7$\%$ relative WER reduction even without using an external LM by replacing its internal LM.
This is more flexible because it does not have additional computational costs by computing an external LM score. If further improvement is needed, shallow fusion can also be used to focus more on linguistic information to improve performance
as shown in the last line of Table~\ref{tab:ls100cross}. In addition, the effect of the Decoupled-AED was consistent on both of these two cross-domain corpora, showing the robust generalisation capability of the proposed method. 

\begin{table}[t] \footnotesize
    \caption{Intra-domain \%WER ($\downarrow$) results obtained on dev/test sets of LibriSpeech for offline/online neural transducer ASR models trained on LS100.}
  \label{tab:ls100-nt}
  \centering
  \setlength{\tabcolsep}{1.2mm}
  \renewcommand\arraystretch{1.37}
  \begin{tabular}{l | c |c| c| c}
    \Xhline{3\arrayrulewidth}
     {\textbf{Offline}}&\multicolumn{2}{c|}{Test}&\multicolumn{2}{c}{Dev}\\
     \cline{2-5}
     Neural Transducer Models&{clean}&{other}&{clean}&{other} \\
    \hline
    w2v2 Transducer\citep{yang2022knowledge}&5.2 &11.8& 5.1&12.2\\
    GM \citep{Ling2021ImprovingPT}&4.3&8.8&4.1&8.8\\
    ATM \citep{Baskar2022Ask2MaskGD}&3.9&8.9&3.7&9.0\\
    \hline
    LSTM T-T Baseline&4.3&7.9&4.1&8.1\\
    Stateless T-T Baseline&4.3&7.6&4.3&7.5\\
    {Transformer T-T} Baseline&\textbf{3.6}&\textbf{6.8}&\textbf{3.5}&\textbf{6.9}\\
    Decoupled-Transducer&3.8&7.1&3.7&7.0\\
    \Xhline{3\arrayrulewidth}
    \textbf{Online} NT Models &--&--&--&--\\
    \hline
    LSTM T-T Baseline&5.3&12.5&5.1&12.5\\
    Stateless T-T Baseline&5.6&12.6&5.5&12.6\\
    Transformer T-T Baseline&5.1&\textbf{12.0}&4.9&\textbf{12.0}\\
    Decoupled-Transducer&\textbf{5.1}&12.2&\textbf{4.9}&12.1\\
    \Xhline{3\arrayrulewidth}
  \end{tabular}
  \vspace{-0.3cm}
\end{table}

\begin{table}[t] \footnotesize
\caption{ASR \%WER ($\downarrow$) results for offline/online Transducer models in cross-domain adaptation scenarios. Internal LM was the replaceable internal LM of the Decoupled-Transducer.}
  \label{tab:ls100cross-nt}
  \centering
  \setlength{\tabcolsep}{1.8mm}
  \renewcommand\arraystretch{1.34}
  \begin{tabular}{l | c c| c c}
    \Xhline{3\arrayrulewidth}
     \textbf{Offline}&\multicolumn{2}{c|}{LS100$\Rightarrow$Ted2}&\multicolumn{2}{c}{LS100$\Rightarrow$AESRC}\\
     Neural Transducer Models&{\;}{Test}&{Dev}{\;}&{\;}{Dev}&{Test}{\;} \\
    \hline
    LSTM T-T Baseline&{\;}11.2&11.9{\;}&{\;}17.7&18.3{\;}\\
    \quad+Target-domain LM SF&{\;}9.9&10.5{\;}&{\;}15.7&16.0{\;}\\
    Stateless T-T Baseline&{\;}10.5&11.3{\;}&{\;}16.5&17.0{\;}\\
    \quad+Target-domain LM SF&{\;}9.8&10.3{\;}&{\;}14.5&15.0{\;}\\
    Transformer T-T Baseline&{\;}10.2&11.1{\;}&{\;}15.7&16.4{\;}\\
    \quad+Target-domain LM SF&{\;}9.6&10.2{\;}&{\;}14.0&14.5{\;}\\
    \cline{1-1}
    Decoupled-Transducer&{\;}10.7&11.5{\;}&{\;}16.3&16.9{\;}\\
    +Replace Internal LM&{\;}\textbf{9.3}&\textbf{10.0}{\;}&{\;}\textbf{14.3}&\textbf{14.7}{\;}\\
    \quad++Target-domain LM SF&{\;}9.0&9.7{\;}&{\;}13.0&13.3{\;}\\
    \Xhline{3\arrayrulewidth}
    \textbf{Online} Neural Transducer&&&&\\
    \hline
    LSTM T-T Baseline&{\;}14.7&14.4{\;}&{\;}28.8&27.5{\;}\\
    \quad+Target-domain LM SF&{\;}13.5&13.2{\;}&{\;}25.8&24.5{\;}\\
    Stateless T-T Baseline&{\;}14.7&14.4{\;}&{\;}28.2&26.9{\;}\\
    \quad+Target-domain LM SF&{\;}12.9&12.9{\;}&{\;}24.9&23.6{\;}\\
    Transformer T-T Baseline&{\;}14.7&14.0{\;}&{\;}27.5&26.3{\;}\\
    \quad+Target-domain LM SF&{\;}13.6&12.9{\;}&{\;}24.7&23.6{\;}\\
    \cline{1-1}
    Decoupled-Transducer&{\;}14.7&14.8{\;}&{\;}28.3&26.9{\;}\\
    +Replace ILM&{\;}\textbf{12.9}&\textbf{12.9}{\;}&{\;}\textbf{25.6}&\textbf{23.9}{\;}\\
    \quad++Target-domain LM SF&{\;}12.2&12.3{\;}&{\;}23.0&21.5{\;}\\
    \Xhline{3\arrayrulewidth}
  \end{tabular}
  \vspace{-0.4cm}
\end{table}
\begin{table}[h] \footnotesize
\caption{ASR \%WER ($\downarrow$) for different online neural transducer methods on intra and cross-domain data. For cross-domain scenarios, the internal LM in HAT \citep{Variani2020HybridAT} was estimated, the one for the factorised T-T \citep{Chen2021FactorizedNT} was fine-tuned, and the one for Decoupled-Transducer was replaced, and shallow fusion was used.}
  \label{tab:compare}
  \centering
  \setlength{\tabcolsep}{1.3mm}
  \renewcommand\arraystretch{1.39}
  \begin{tabular}{l | c c| c| c}
    \Xhline{3\arrayrulewidth}
     \hspace{-1.0mm}\textbf{Online}&\multicolumn{2}{c|}{LS100 Test}&{Ted2}&{AESRC}\\
     \hspace{-1.0mm}Neural Transducer Models&{clean}&{other}&{Test}&{Test} \\
    \hline
    \hspace{-1.0mm}Transformer T-T Baseline&5.1&\textbf{12.0}&13.6&23.6\\
    \hspace{-1.0mm}HAT \citep{Variani2020HybridAT}&5.4&12.2&13.6&23.0\\
    \hspace{-1.0mm}Factorised T-T\citep{Chen2021FactorizedNT}&5.4&12.4&13.3&22.5\\
    \hline
    \hspace{-1.0mm}Decoupled-Transducer&\textbf{5.1}&12.2&\textbf{12.2}&\textbf{21.5}\\
    \Xhline{3\arrayrulewidth}
  \end{tabular}
  \vspace{-0.4cm}
\end{table}
\begin{table*}[t] \footnotesize
    \caption{Ablation studies on the effects of internal LM logits in intra-domain and cross-domain scenarios.
    Intra-domain WER ($\downarrow$) results were obtained on dev/test sets of LibriSpeech for models trained on the 100-hour subset (LS100), while cross-domain performance was evaluated on TED-LIUM 2 (Ted2) and AESRC2020. The decoding logits refer to the logits used during decoding, while the source and target ${\rm logits}^{\rm LM}$ respectively denote the logits output by source and target internal LMs.}
    \label{ablation}
  \centering
  \setlength{\tabcolsep}{2.2mm}
  \renewcommand\arraystretch{1.4}
  \begin{tabular}{l  c |c c c c|c c|c c}
    \Xhline{3\arrayrulewidth}
     \multirow{2}{*}{ASR Models}&\multirow{2}{*}{Decoding Logits}&\multicolumn{2}{c}{LS100 Test}&\multicolumn{2}{c|}{LS100 Dev}&\multicolumn{2}{c|}{LS100$\Rightarrow$Ted2}&\multicolumn{2}{c}{LS100$\Rightarrow$AESRC2020}\\
     &&{clean}&{other}&{clean}&{other}&{~}Test&{Dev}&{\quad\ }Dev&Test \\
    \hline
    Decoupled-AED&${\rm logits}^{\rm AC}$&3.8&7.0&3.6&7.0&{~}10.1&10.8&{\quad\ }15.4&16.2\\
    Decoupled-AED&${\rm logits}^{\rm AC}$ w/ source ${\rm logits}^{\rm LM}$&3.4&6.4&3.3&6.4&{~}9.8&10.5&{\quad\ }14.8&15.5\\
    Decoupled-AED&${\rm logits}^{\rm AC}$ w/ target ${\rm logits}^{\rm LM}$&--&--&--&--&{~}9.0&9.6&{\quad\ }13.5&14.0\\
    \hline
    Online Decoupled-Transducer&${\rm logits}^{\rm AC}$&5.7&13.3&5.7&13.2&{~}15.1&15.0&{\quad\ }30.2&28.6\\
    Online Decoupled-Transducer&${\rm logits}^{\rm AC}$ w/ source ${\rm logits}^{\rm LM}$&5.1&12.2&4.9&12.1&{~}14.7&14.8&{\quad\ }28.3&26.9\\
    Online Decoupled-Transducer&${\rm logits}^{\rm AC}$ w/ target ${\rm logits}^{\rm LM}$&--&--&--&--&{~}12.9&12.9&{\quad\ }25.6&23.9\\
    \Xhline{3\arrayrulewidth}
  \end{tabular}
  \vspace{-0.4cm}
\end{table*}
\vspace{-0.1cm}
\subsection{Experiments on Decoupled-Transducer}
Table~\ref{tab:ls100-nt} lists the intra-domain results for offline/online neural transducer models and showed that our transducer models achieved good results on the LS100 benchmark. In addition, the LSTM T-T and Stateless T-T showed similar performance to each other in both offline and online scenarios, which is consistent with the conclusion of \citep{9054419}. 
\vspace{-0.01cm}
However, the Transformer T-T greatly outperformed the other two baseline models, indicating that the Transformer prediction network was still effective to achieve further performance improvement. Furthermore, the proposed Decoupled-Transducer still achieved competitive results to the strong Transformer T-T baseline model.

Experiments were then conducted to compare the cross-domain ASR performance for offline/online transducer models on the TED-LIUM 2 and AESRC2020 corpora. As shown in Table~\ref{tab:ls100cross-nt},
the Transformer T-T baseline outperformed the other two baseline models in the cross-domain scenario also and the cross-domain ASR accuracy could be further improved with the help of external target-domain LM via shallow fusion. However, without the external LM,
the proposed Decoupled-Transducer with the internal LM replaced already performs virtually as
well as the strong Transformer T-T with shallow fusion. When shallow fusion was also used for the Decoupled-Transducer, up to 8.3$\%$ and 10.3$\%$ relative WER reduction in offline and online scenarios compared with the best results of the baseline models were obtained. In addition, the Decoupled-Transducer was shown to be effective for both offline and online transducer models on both cross-domain corpora showing consistent strong performance on domain adaptation.
\vspace{-0.1cm}
\subsubsection{Comparison with Related Work}
We also implemented 
HAT \citep{Variani2020HybridAT} 
and factorised T-T \citep{Chen2021FactorizedNT} 
based on the Transformer T-T baseline model to compare with our proposed decoupled structure.
Table~\ref{tab:compare} shows that 
HAT (371M parameters) and factorised T-T (372M parameters) slightly
degraded intra-domain performance compared to strong Transformer T-T (371M parameters). Nevertheless, leveraging their advantages in domain adaptation (i.e., internal estimation or adaptation) compensates for this issue, leading to superior performance over Transformer T-T in cross-domain scenarios. However,
the Decoupled-Transducer (372M parameters) still surpassed HAT and 
factorised T-T in both intra and cross-domain scenarios. 
The internal LM estimation in HAT complicates decoding and may not always be accurate \citep{tsunoo22_interspeech}. In addition, the neural transducer loss, which permits multiple non-blank outputs at a single time step \citep{Graves2012SequenceTW}, can potentially present convergence difficulties for factorised T-T, which directly adds encoder output logits to internal LM log probabilities lacking dynamic weights. Moreover, the direct replacement of the internal LM in the Decoupled-Transducer is more flexible.


To conclude, the proposed decoupled structure was shown to be consistently effective for both AED and neural transducer models on ASR domain adaptation while retaining intra-domain accuracy.
\subsection{Ablation Studies}
In this section, we report ablation studies on the effects of the internal LM logits. In the proposed decoupled structure, the acoustic and linguistic information are jointly learned via logit addition, which is modular and flexible to validate the effects of the logits obtained from linguistic information. The results are shown in Table~\ref{ablation}, where using ${\rm logits}^{\rm AC}$ as the decoding logits means directly utilising the output of the acoustic part for decoding without internal LM prediction, ${\rm logits}^{\rm AC}$ with source ${\rm logits}^{\rm LM}$ stood for the Decoupled-AED or Decoupled-Transducer in the normal case,
and ${\rm logits}^{\rm AC}$ with target ${\rm logits}^{\rm LM}$ represented the decoupled structure with the internal LM replaced by a target-domain LM.

Intuitively, removing the logits of the source-domain internal LM would lead to performance degradation in both intra-domain and cross-domain scenarios, which shows the importance of linguistic information. However, the improvement brought by the source-domain internal LM is weakened in cross-domain scenarios compared to the source-domain scenarios. To be more specific, for the Decoupled-AED,  having the source ${\rm logits}^{\rm LM}$ could yield around 10$\%$ relative WER reduction in source-domain test sets compared to only using the ${\rm logits}^{\rm AC}$, but the improvement dropped to around 3$\%$ relative in cross-domain TED-LIUM 2 data and around 4$\%$ relative in cross-domain AESRC2020 data. This effect for the online Decoupled-Transducer
was similar. This was caused by the domain mismatch because the source-domain internal LM learned a different data distribution to the target domain. However, when the internal LM was replaced, significant cross-domain improvements were obtained, with up to 13.6$\%$ relative WER reduction
for the Decoupled-AED and 16.4$\%$ for online Decoupled-Transducer compared to only using the ${\rm logits}^{\rm AC}$.


The WER improvement brought by the target-domain internal LM over the source-domain internal LM on both TED-LIUM 2 and AESRC2020 test sets is statistically significant at the 0.1\% level using the matched-pair sentence-segment word error statistical test \citep{115546}.

Therefore, it can be concluded that the linguistic knowledge learned by the internal LM plays an important role in ASR performance but is at risk of domain mismatch, which can be effectively resolved by replacing it with a target-domain LM under the flexible decoupled structure.
\vspace{-0.05cm}
\subsection{Application to E2E Speech Translation}
The proposed Decoupled-AED structure was also applied to
the domain adaptation for E2E speech translation.
The proposed Decoupled-AED is very similar when applied to either ASR or ST tasks, the only difference is that the translation is used as the target text instead of the transcript\footnote{Note that CTC/attention joint translation \citep{Deng2022BlockwiseST} is employed for the ST task. This means the target language translation is directly used to supervise the CTC branch which has reordering capability with the Transformer global attention \citep{DBLP:conf/acl/2021f}. During joint translation, CTC prediction for translation is also considered and the specific implementation is similar to ASR joint decoding. The CTC weight during translation was 0.7 and $\beta$ in Eq.~\ref{logits} was 1.}. E2E ST models were trained on the Fisher-CallHome Spanish\footnote{Fisher-CallHome Spanish is a Spanish (ES) to English (EN) ST corpus and includes spontaneous conversations between friends and family.} \citep{post-etal-2013-improved} following the ESPnet \citep{Watanabe2018ESPnet} recipe and evaluated on the Europarl-ST \citep{jairsan2020a} Spanish-English language pair collected from the European Parliament. 
The multilingual wav2vec2.0 (XLSR-53) encoder \citep{conneau21_interspeech} provided by Fairseq (i.e. "xlsr\_53\_56k") was used as the encoder, which was pre-trained first on the ASR task before used in the ST task to achieve better performance following \cite{inaguma-etal-2020-espnet}.

\begin{table}[t] \footnotesize
\vspace{-0.1cm}
\caption{Intra-domain \%BLEU ($\uparrow$) results obtained on test sets of Fisher-CallHome Spanish for AED ST models. Transformer is abbreviated as Trans.
The devtest and evltest sets of CallHome were abbreviated as dev and evl. Case-insensitive BLEU was reported on Fisher-\{dev, dev2, test\} (with 4 references), and CallHome-\{devtest, evltest\} (with single reference).}
  \label{tab:st}
  \centering
  \setlength{\tabcolsep}{1.3mm}
  \renewcommand\arraystretch{1.4}
  \begin{tabular}{l | c c c |c c}
    \Xhline{3\arrayrulewidth}
     \hspace{-1.0mm}\multirow{2}{*}{AED Models}&\multicolumn{3}{c|}{Fisher}&\multicolumn{2}{c}{Callhome}\\
     &{dev}&{dev2}&{test}&{dev}&evl \\
    \hline
    \hspace{-1.0mm}Cascade \citep{dalmia-etal-2021-searchable}&50.4 &51.2 &50.7& 19.6& 19.2\\
    \hspace{-1.0mm}ESPnet \citep{inaguma-etal-2020-espnet}&48.9 &49.3& 48.4& 18.8& 18.7\\
    \hspace{-1.0mm}Trans. MD\citep{dalmia-etal-2021-searchable}&55.2 &55.2& 55.0& 21.7&21.5\\ 
    \hspace{-1.0mm}Fast MD \citep{9687894}&54.8 &55.1& 54.4& 21.3& 21.3\\
    \hline
    \hspace{-1.0mm}Normal AED Baseline&56.1&\textbf{56.7}&\textbf{55.5}&24.5&24.0\\
    \hspace{-1.0mm}Decoupled-AED&\textbf{56.1}&56.6&55.3&\textbf{24.6}&\textbf{24.4}\\
    \Xhline{3\arrayrulewidth}
  \end{tabular}
\end{table}
\begin{table}[t] \footnotesize
\caption{ST \%BLEU ($\uparrow$) results in cross-domain adaptation scenario. Fisher-CallHome Spanish corpus is denoted as FCHS and Europarl-ST refers to the Spanish-English direction. Case-insensitive BLEU was reported. LM is a target-language LM.}
  \label{tab:stcross}
  \centering
  \setlength{\tabcolsep}{1.8mm}
  \renewcommand\arraystretch{1.4}
  \begin{tabular}{l | c c}
    \Xhline{3\arrayrulewidth}
     \multirow{2}{*}{AED Models}&\multicolumn{2}{c}{FCHS$\Rightarrow$Europarl-ST}\\
     &{\quad\,}{Test}&{Dev} \\
    \hline
    Normal AED Baseline&{\quad\,}12.5&13.8\\
    \quad+Target-domain LM shallow fusion&{\quad\,}13.4&15.1\\
    Decoupled-AED&{\quad\,}12.4&13.9\\
    {\ }+Replace Internal LM&{\quad\,}\textbf{13.6}&\textbf{15.1}\\
    \quad++Target-domain LM shallow fusion&{\quad\,}14.0&15.6\\
    \Xhline{3\arrayrulewidth}
  \end{tabular}
  \vspace{-0.25cm}
\end{table}

The intra-domain ST BLEU \citep{papineni2002bleu} results are listed in Table~\ref{tab:st}, which shows that our ST models surpassed previous systems on the Fisher-CallHome Spanish benchmark. 
Compared to the strong AED baseline,
the proposed Decoupled-AED model still achieved similar ST performance 
for the intra-domain scenario, which showed that decoupling the AED decoder into
the acoustic and linguistic component parts also did not degrade performance for intra-domain ST.

Experiments were then conducted to compare the cross-domain ST performance on the Europarl-ST Spanish-English data and the results are
shown in Table~\ref{tab:stcross}. 
The proposed decoupled structure with the internal LM replaced by a target-domain target-language LM outperformed the baseline model by $+1.3$ BLEU points in cross-domain performance.
As found for the ASR task, the AED baseline model improves the cross-domain performance by including an external LM via shallow fusion and achieved results close to the Decoupled-AED without an external LM. Note that the Decoupled-AED does not complicate the decoding process.
Furthermore, when shallow fusion was also employed for the decoupled structure, better cross-domain translation quality could be achieved.
\section{Conclusion}
\label{conclusion}
This paper proposes a decoupled structure for E2E ASR models to achieve flexible domain adaptation and applies it to two E2E ASR models: the attention-based encoder-decoder and the neural transducer. In the proposed decoupled structure, the acoustic and linguistic parts of the E2E model decoder/prediction network are separated, making the linguistic part (i.e. the internal LM) replaceable. When encountering a domain shift, the internal LM can be directly replaced by a target-domain LM and thus be flexibly adapted to the target domain. The final logits of the decoupled structure are the sum of the logits computed from acoustic and linguistic information, making the prediction more modular. 
Experiments showed that the decoupled structure achieved up to 15.1$\%$ and 17.2$\%$ relative WER reduction on TED-LIUM 2 and AESRC2020 cross-domain corpora
while maintaining intra-domain results. It was also shown that the decoupled structure could also be used to boost cross-domain speech translation quality while retaining the intra-domain performance.
\section*{Acknowledgement}
Keqi Deng is funded by the Cambridge Trust. This work has been performed using resources provided by the Cambridge Tier-2 system operated by the University of Cambridge Research Computing Service (www.hpc.cam.ac.uk) funded by EPSRC Tier-2 capital grant EP/T022159/1.




\printcredits

\bibliographystyle{cas-model2-names}

\bibliography{cas-refs}

\begin{thebibliography}{69}
\expandafter\ifx\csname natexlab\endcsname\relax\def\natexlab#1{#1}\fi
\providecommand{\url}[1]{\texttt{#1}}
\providecommand{\href}[2]{#2}
\providecommand{\path}[1]{#1}
\providecommand{\DOIprefix}{doi:}
\providecommand{\ArXivprefix}{arXiv:}
\providecommand{\URLprefix}{URL: }
\providecommand{\Pubmedprefix}{pmid:}
\providecommand{\doi}[1]{\href{http://dx.doi.org/#1}{\path{#1}}}
\providecommand{\Pubmed}[1]{\href{pmid:#1}{\path{#1}}}
\providecommand{\bibinfo}[2]{#2}
\ifx\xfnm\relax \def\xfnm[#1]{\unskip,\space#1}\fi
\bibitem[{Ao et~al.(2021)Ao, Wang, Zhou, Liu, Ren, Wu, Ko, Li, Zhang, Wei,
  Qian, Li and Wei}]{Ao2021SpeechT5UE}
\bibinfo{author}{Ao, J.}, \bibinfo{author}{Wang, R.}, \bibinfo{author}{Zhou,
  L.}, \bibinfo{author}{Liu, S.}, \bibinfo{author}{Ren, S.},
  \bibinfo{author}{Wu, Y.}, \bibinfo{author}{Ko, T.}, \bibinfo{author}{Li, Q.},
  \bibinfo{author}{Zhang, Y.}, \bibinfo{author}{Wei, Z.},
  \bibinfo{author}{Qian, Y.}, \bibinfo{author}{Li, J.}, \bibinfo{author}{Wei,
  F.}, \bibinfo{year}{2021}.
\newblock \bibinfo{title}{{SpeechT5}: Unified-modal encoder-decoder
  pre-training for spoken language processing}, in: \bibinfo{booktitle}{Proc.
  {ACL}}, \bibinfo{address}{Dublin, Ireland}.
\bibitem[{Ao et~al.(2022)Ao, Zhang, Zhou, Liu, Li, Ko, Dai, Li, Qian and
  Wei}]{ao22_interspeech}
\bibinfo{author}{Ao, J.}, \bibinfo{author}{Zhang, Z.}, \bibinfo{author}{Zhou,
  L.}, \bibinfo{author}{Liu, S.}, \bibinfo{author}{Li, H.},
  \bibinfo{author}{Ko, T.}, \bibinfo{author}{Dai, L.}, \bibinfo{author}{Li,
  J.}, \bibinfo{author}{Qian, Y.}, \bibinfo{author}{Wei, F.},
  \bibinfo{year}{2022}.
\newblock \bibinfo{title}{Pre-training transformer decoder for end-to-end {ASR}
  model with unpaired speech data}, in: \bibinfo{booktitle}{Proc. Interspeech},
  \bibinfo{address}{Incheon, Korea}.
\bibitem[{Baskar et~al.(2022)Baskar, Rosenberg, Ramabhadran, Zhang and
  Moreno}]{Baskar2022Ask2MaskGD}
\bibinfo{author}{Baskar, M.K.}, \bibinfo{author}{Rosenberg, A.},
  \bibinfo{author}{Ramabhadran, B.}, \bibinfo{author}{Zhang, Y.},
  \bibinfo{author}{Moreno, P.J.}, \bibinfo{year}{2022}.
\newblock \bibinfo{title}{{Ask2Mask}: Guided data selection for masked speech
  modeling}.
\newblock \bibinfo{journal}{IEEE Journal of Selected Topics in Signal
  Processing} \bibinfo{volume}{16}, \bibinfo{pages}{1357--1366}.
\bibitem[{Boyer et~al.(2021)Boyer, Shinohara, Ishii, Inaguma and
  Watanabe}]{9688251}
\bibinfo{author}{Boyer, F.}, \bibinfo{author}{Shinohara, Y.},
  \bibinfo{author}{Ishii, T.}, \bibinfo{author}{Inaguma, H.},
  \bibinfo{author}{Watanabe, S.}, \bibinfo{year}{2021}.
\newblock \bibinfo{title}{A study of transducer based end-to-end asr with
  espnet: Architecture, auxiliary loss and decoding strategies}, in:
  \bibinfo{booktitle}{Proc. ASRU}, \bibinfo{address}{Cartagena, Colombia}.
\bibitem[{Cao et~al.(2021)Cao, Kang, Fu, Xu, Sun, Zhang and
  Ma}]{Cao2021ImprovingST}
\bibinfo{author}{Cao, S.}, \bibinfo{author}{Kang, Y.}, \bibinfo{author}{Fu,
  Y.}, \bibinfo{author}{Xu, X.}, \bibinfo{author}{Sun, S.},
  \bibinfo{author}{Zhang, Y.}, \bibinfo{author}{Ma, L.}, \bibinfo{year}{2021}.
\newblock \bibinfo{title}{Improving streaming transformer based {ASR} under a
  framework of self-supervised learning}, in: \bibinfo{booktitle}{Proc.
  Interspeech}, \bibinfo{address}{Brno, Czechia}.
\bibitem[{Chan et~al.(2016)Chan, Jaitly, Le and Vinyals}]{Chan2016ListenAA}
\bibinfo{author}{Chan, W.}, \bibinfo{author}{Jaitly, N.}, \bibinfo{author}{Le,
  Q.V.}, \bibinfo{author}{Vinyals, O.}, \bibinfo{year}{2016}.
\newblock \bibinfo{title}{Listen, attend and spell: A neural network for large
  vocabulary conversational speech recognition}, in: \bibinfo{booktitle}{Proc.
  ICASSP}, \bibinfo{address}{Shanghai, China}.
\bibitem[{Chen et~al.(2022)Chen, Meng, Parthasarathy and
  Li}]{Chen2021FactorizedNT}
\bibinfo{author}{Chen, X.}, \bibinfo{author}{Meng, Z.},
  \bibinfo{author}{Parthasarathy, S.}, \bibinfo{author}{Li, J.},
  \bibinfo{year}{2022}.
\newblock \bibinfo{title}{Factorized neural transducer for efficient language
  model adaptation}, in: \bibinfo{booktitle}{Proc. ICASSP},
  \bibinfo{address}{Singapore}.
\bibitem[{Chorowski et~al.(2015)Chorowski, Bahdanau, Serdyuk, Cho and
  Bengio}]{chorowski2015attention}
\bibinfo{author}{Chorowski, J.K.}, \bibinfo{author}{Bahdanau, D.},
  \bibinfo{author}{Serdyuk, D.}, \bibinfo{author}{Cho, K.},
  \bibinfo{author}{Bengio, Y.}, \bibinfo{year}{2015}.
\newblock \bibinfo{title}{Attention-based models for speech recognition}, in:
  \bibinfo{booktitle}{Proc. NeurIPS}, \bibinfo{address}{Montreal, Canada}.
\bibitem[{Choudhury et~al.(2022)Choudhury, Gandhe, Ding and Bulyko}]{9746480}
\bibinfo{author}{Choudhury, C.}, \bibinfo{author}{Gandhe, A.},
  \bibinfo{author}{Ding, X.}, \bibinfo{author}{Bulyko, I.},
  \bibinfo{year}{2022}.
\newblock \bibinfo{title}{{A likelihood ratio based domain adaptation method
  for E2E models}}, in: \bibinfo{booktitle}{Proc. ICASSP},
  \bibinfo{address}{Singapore}.
\bibitem[{Chuang et~al.(2021)Chuang, Chuang, Chang and
  Lee}]{DBLP:conf/acl/2021f}
\bibinfo{author}{Chuang, S.}, \bibinfo{author}{Chuang, Y.},
  \bibinfo{author}{Chang, C.}, \bibinfo{author}{Lee, H.}, \bibinfo{year}{2021}.
\newblock \bibinfo{title}{Investigating the reordering capability in ctc-based
  non-autoregressive end-to-end speech translation}, in:
  \bibinfo{booktitle}{{ACL/IJCNLP} (Findings)}, \bibinfo{address}{Bangkok,
  Thailand}.
\bibitem[{Conneau et~al.(2021)Conneau, Baevski, Collobert, Mohamed and
  Auli}]{conneau21_interspeech}
\bibinfo{author}{Conneau, A.}, \bibinfo{author}{Baevski, A.},
  \bibinfo{author}{Collobert, R.}, \bibinfo{author}{Mohamed, A.},
  \bibinfo{author}{Auli, M.}, \bibinfo{year}{2021}.
\newblock \bibinfo{title}{Unsupervised cross-lingual representation learning
  for speech recognition}, in: \bibinfo{booktitle}{Proc. Interspeech},
  \bibinfo{address}{Brno, Czechia}.
\bibitem[{Dahl et~al.(2012)Dahl, Yu, Deng and Acero}]{5740583}
\bibinfo{author}{Dahl, G.E.}, \bibinfo{author}{Yu, D.}, \bibinfo{author}{Deng,
  L.}, \bibinfo{author}{Acero, A.}, \bibinfo{year}{2012}.
\newblock \bibinfo{title}{Context-dependent pre-trained deep neural networks
  for large-vocabulary speech recognition}.
\newblock \bibinfo{journal}{IEEE Transactions on Audio, Speech, and Language
  Processing} \bibinfo{volume}{20}, \bibinfo{pages}{30--42}.
\bibitem[{Dalmia et~al.(2021)Dalmia, Yan, Raunak, Metze and
  Watanabe}]{dalmia-etal-2021-searchable}
\bibinfo{author}{Dalmia, S.}, \bibinfo{author}{Yan, B.},
  \bibinfo{author}{Raunak, V.}, \bibinfo{author}{Metze, F.},
  \bibinfo{author}{Watanabe, S.}, \bibinfo{year}{2021}.
\newblock \bibinfo{title}{Searchable hidden intermediates for end-to-end models
  of decomposable sequence tasks}, in: \bibinfo{booktitle}{{NAACL-HLT}},
  \bibinfo{address}{Mexico City, Mexico}.
\bibitem[{Deng et~al.(2021)Deng, Cao, Zhang and Ma}]{9688009}
\bibinfo{author}{Deng, K.}, \bibinfo{author}{Cao, S.}, \bibinfo{author}{Zhang,
  Y.}, \bibinfo{author}{Ma, L.}, \bibinfo{year}{2021}.
\newblock \bibinfo{title}{Improving hybrid {CTC}/attention end-to-end speech
  recognition with pretrained acoustic and language models}, in:
  \bibinfo{booktitle}{Proc. ASRU}, \bibinfo{address}{Cartagena, Colombia}.
\bibitem[{Deng et~al.(2022)Deng, Watanabe, Shi and Arora}]{Deng2022BlockwiseST}
\bibinfo{author}{Deng, K.}, \bibinfo{author}{Watanabe, S.},
  \bibinfo{author}{Shi, J.}, \bibinfo{author}{Arora, S.}, \bibinfo{year}{2022}.
\newblock \bibinfo{title}{Blockwise streaming transformer for spoken language
  understanding and simultaneous speech translation}, in:
  \bibinfo{booktitle}{Proc. Interspeech}, \bibinfo{address}{Incheon, Korea}.
\bibitem[{Deng and Woodland(2023)}]{keqi2023rilm}
\bibinfo{author}{Deng, K.}, \bibinfo{author}{Woodland, P.C.},
  \bibinfo{year}{2023}.
\newblock \bibinfo{title}{{Adaptable end-to-end ASR models using replaceable
  internal LMs and residual softmax}}, in: \bibinfo{booktitle}{Proc. ICASSP},
  \bibinfo{address}{Rhodes Island, Greece}.
\bibitem[{Du et~al.(2022)Du, Wang, Zhang, Chen, Xu, Xie and
  Chen}]{Non-Parametric2022}
\bibinfo{author}{Du, Y.}, \bibinfo{author}{Wang, W.}, \bibinfo{author}{Zhang,
  Z.}, \bibinfo{author}{Chen, B.}, \bibinfo{author}{Xu, T.},
  \bibinfo{author}{Xie, J.}, \bibinfo{author}{Chen, E.}, \bibinfo{year}{2022}.
\newblock \bibinfo{title}{Non-parametric domain adaptation for end-to-end
  speech translation}, in: \bibinfo{booktitle}{{EMNLP}}, \bibinfo{address}{Abu
  Dhabi, United Arab Emirates}.
\bibitem[{Fu et~al.(2022)Fu, Li, Li, Deng, Li, Fan, Chen and He}]{Fu2022UFO2AU}
\bibinfo{author}{Fu, L.}, \bibinfo{author}{Li, S.}, \bibinfo{author}{Li, Q.},
  \bibinfo{author}{Deng, L.}, \bibinfo{author}{Li, F.}, \bibinfo{author}{Fan,
  L.}, \bibinfo{author}{Chen, M.}, \bibinfo{author}{He, X.},
  \bibinfo{year}{2022}.
\newblock \bibinfo{title}{{UFO2}: A unified pre-training framework for online
  and offline speech recognition}.
\newblock \bibinfo{journal}{ArXiv} \bibinfo{volume}{abs/2210.14515}.
\bibitem[{Gage(1994)}]{gage1994}
\bibinfo{author}{Gage, P.}, \bibinfo{year}{1994}.
\newblock \bibinfo{title}{A new algorithm for data compression}.
\newblock \bibinfo{journal}{The C Users Journal} \bibinfo{volume}{12},
  \bibinfo{pages}{23--38}.
\bibitem[{Ghodsi et~al.(2020)Ghodsi, Liu, Apfel, Cabrera and
  Weinstein}]{9054419}
\bibinfo{author}{Ghodsi, M.}, \bibinfo{author}{Liu, X.},
  \bibinfo{author}{Apfel, J.}, \bibinfo{author}{Cabrera, R.},
  \bibinfo{author}{Weinstein, E.}, \bibinfo{year}{2020}.
\newblock \bibinfo{title}{{RNN}-transducer with stateless prediction network},
  in: \bibinfo{booktitle}{Proc. ICASSP}, \bibinfo{address}{Barcelona, Spain}.
\bibitem[{Graves(2012)}]{Graves2012SequenceTW}
\bibinfo{author}{Graves, A.}, \bibinfo{year}{2012}.
\newblock \bibinfo{title}{Sequence transduction with recurrent neural
  networks}.
\newblock \bibinfo{journal}{ArXiv} \bibinfo{volume}{abs/1211.3711}.
\bibitem[{Graves et~al.(2006)Graves, Fern{\'a}ndez, Gomez and
  Schmidhuber}]{graves2006connectionist}
\bibinfo{author}{Graves, A.}, \bibinfo{author}{Fern{\'a}ndez, S.},
  \bibinfo{author}{Gomez, F.}, \bibinfo{author}{Schmidhuber, J.},
  \bibinfo{year}{2006}.
\newblock \bibinfo{title}{Connectionist temporal classification: Labelling
  unsegmented sequence data with recurrent neural networks}, in:
  \bibinfo{booktitle}{Proc. {ICML}}, \bibinfo{address}{Pittsburgh,
  Pennsylvania, USA}.
\bibitem[{Graves and Jaitly(2014)}]{10.5555/3044805.3045089}
\bibinfo{author}{Graves, A.}, \bibinfo{author}{Jaitly, N.},
  \bibinfo{year}{2014}.
\newblock \bibinfo{title}{Towards end-to-end speech recognition with recurrent
  neural networks}, in: \bibinfo{booktitle}{Proc. ICML},
  \bibinfo{address}{Beijing, China}.
\bibitem[{Graves et~al.(2013)Graves, Mohamed and Hinton}]{6638947}
\bibinfo{author}{Graves, A.}, \bibinfo{author}{Mohamed, A.},
  \bibinfo{author}{Hinton, G.}, \bibinfo{year}{2013}.
\newblock \bibinfo{title}{Speech recognition with deep recurrent neural
  networks}, in: \bibinfo{booktitle}{Proc. ICASSP},
  \bibinfo{address}{Vancouver, BC, Canada}.
\bibitem[{Gulati et~al.(2020)Gulati, Qin, Chiu, Parmar, Zhang, Yu, Han, Wang,
  Zhang, Wu and Pang}]{gulati20_interspeech}
\bibinfo{author}{Gulati, A.}, \bibinfo{author}{Qin, J.}, \bibinfo{author}{Chiu,
  C.C.}, \bibinfo{author}{Parmar, N.}, \bibinfo{author}{Zhang, Y.},
  \bibinfo{author}{Yu, J.}, \bibinfo{author}{Han, W.}, \bibinfo{author}{Wang,
  S.}, \bibinfo{author}{Zhang, Z.}, \bibinfo{author}{Wu, Y.},
  \bibinfo{author}{Pang, R.}, \bibinfo{year}{2020}.
\newblock \bibinfo{title}{Conformer: Convolution-augmented transformer for
  speech recognition}, in: \bibinfo{booktitle}{Proc. Interspeech},
  \bibinfo{address}{Shanghai, China}.
\bibitem[{Gulcchre et~al.(2015)Gulcchre, Firat, Xu, Cho, Barrault, Lin,
  Bougares, Schwenk and Bengio}]{gulcehre2015using}
\bibinfo{author}{Gulcchre, C.}, \bibinfo{author}{Firat, O.},
  \bibinfo{author}{Xu, K.}, \bibinfo{author}{Cho, K.},
  \bibinfo{author}{Barrault, L.}, \bibinfo{author}{Lin, H.},
  \bibinfo{author}{Bougares, F.}, \bibinfo{author}{Schwenk, H.},
  \bibinfo{author}{Bengio, Y.}, \bibinfo{year}{2015}.
\newblock \bibinfo{title}{On using monolingual corpora in neural machine
  translation}.
\newblock \bibinfo{journal}{arXiv preprint arXiv:1503.03535} .
\bibitem[{Higuchi et~al.(2022)Higuchi, Yan, Arora, Ogawa, Kobayashi and
  Watanabe}]{Higuchi2022BERTMC}
\bibinfo{author}{Higuchi, Y.}, \bibinfo{author}{Yan, B.},
  \bibinfo{author}{Arora, S.}, \bibinfo{author}{Ogawa, T.},
  \bibinfo{author}{Kobayashi, T.}, \bibinfo{author}{Watanabe, S.},
  \bibinfo{year}{2022}.
\newblock \bibinfo{title}{{BERT} meets {CTC}: New formulation of end-to-end
  speech recognition with pre-trained masked language model}, in:
  \bibinfo{booktitle}{Proc. {EMNLP} (Findings)}, \bibinfo{address}{Abu Dhabi,
  United Arab Emirates}.
\bibitem[{Hinton et~al.(2012)Hinton, Deng, Yu, Dahl, rahman Mohamed, Jaitly,
  Senior, Vanhoucke, Nguyen, Sainath and Kingsbury}]{Hinton2012DeepNN}
\bibinfo{author}{Hinton, G.E.}, \bibinfo{author}{Deng, L.},
  \bibinfo{author}{Yu, D.}, \bibinfo{author}{Dahl, G.E.},
  \bibinfo{author}{rahman Mohamed, A.}, \bibinfo{author}{Jaitly, N.},
  \bibinfo{author}{Senior, A.W.}, \bibinfo{author}{Vanhoucke, V.},
  \bibinfo{author}{Nguyen, P.}, \bibinfo{author}{Sainath, T.N.},
  \bibinfo{author}{Kingsbury, B.}, \bibinfo{year}{2012}.
\newblock \bibinfo{title}{Deep neural networks for acoustic modeling in speech
  recognition}.
\newblock \bibinfo{journal}{IEEE Signal Processing Magazine}
  \bibinfo{volume}{29}, \bibinfo{pages}{82}.
\bibitem[{Hochreiter and Schmidhuber(1997)}]{Hochreiter1997}
\bibinfo{author}{Hochreiter, S.}, \bibinfo{author}{Schmidhuber, J.},
  \bibinfo{year}{1997}.
\newblock \bibinfo{title}{Long short-term memory}.
\newblock \bibinfo{journal}{Neural Computation} \bibinfo{volume}{9},
  \bibinfo{pages}{1735--1780}.
\bibitem[{Hsu et~al.(2021)Hsu, Sriram, Baevski, Likhomanenko, Xu, Pratap, Kahn,
  Lee, Collobert, Synnaeve and Auli}]{hsu21_interspeech}
\bibinfo{author}{Hsu, W.}, \bibinfo{author}{Sriram, A.},
  \bibinfo{author}{Baevski, A.}, \bibinfo{author}{Likhomanenko, T.},
  \bibinfo{author}{Xu, Q.}, \bibinfo{author}{Pratap, V.},
  \bibinfo{author}{Kahn, J.}, \bibinfo{author}{Lee, A.},
  \bibinfo{author}{Collobert, R.}, \bibinfo{author}{Synnaeve, G.},
  \bibinfo{author}{Auli, M.}, \bibinfo{year}{2021}.
\newblock \bibinfo{title}{{Robust wav2vec 2.0: Analyzing domain shift in
  self-supervised pre-training}}, in: \bibinfo{booktitle}{Proc. Interspeech},
  \bibinfo{address}{Brno, Czechia}.
\bibitem[{Inaguma et~al.(2021)Inaguma, Dalmia, Yan and Watanabe}]{9687894}
\bibinfo{author}{Inaguma, H.}, \bibinfo{author}{Dalmia, S.},
  \bibinfo{author}{Yan, B.}, \bibinfo{author}{Watanabe, S.},
  \bibinfo{year}{2021}.
\newblock \bibinfo{title}{{Fast-MD}: Fast multi-decoder end-to-end speech
  translation with non-autoregressive hidden intermediates}, in:
  \bibinfo{booktitle}{Proc. ASRU}, \bibinfo{address}{Cartagena, Colombia}.
\bibitem[{Inaguma et~al.(2020)Inaguma, Kiyono, Duh, Karita, Yalta, Hayashi and
  Watanabe}]{inaguma-etal-2020-espnet}
\bibinfo{author}{Inaguma, H.}, \bibinfo{author}{Kiyono, S.},
  \bibinfo{author}{Duh, K.}, \bibinfo{author}{Karita, S.},
  \bibinfo{author}{Yalta, N.}, \bibinfo{author}{Hayashi, T.},
  \bibinfo{author}{Watanabe, S.}, \bibinfo{year}{2020}.
\newblock \bibinfo{title}{{ESP}net-{ST}: All-in-one speech translation
  toolkit}, in: \bibinfo{booktitle}{Proc. {ACL} (demo)},
  \bibinfo{address}{Seattle, Washington, USA}.
\bibitem[{{Iranzo-Sánchez} et~al.(2020){Iranzo-Sánchez}, {Silvestre-Cerdà},
  {Jorge}, {Roselló}, {Giménez}, {Sanchis}, {Civera} and
  {Juan}}]{jairsan2020a}
\bibinfo{author}{{Iranzo-Sánchez}, J.}, \bibinfo{author}{{Silvestre-Cerdà},
  J.A.}, \bibinfo{author}{{Jorge}, J.}, \bibinfo{author}{{Roselló}, N.},
  \bibinfo{author}{{Giménez}, A.}, \bibinfo{author}{{Sanchis}, A.},
  \bibinfo{author}{{Civera}, J.}, \bibinfo{author}{{Juan}, A.},
  \bibinfo{year}{2020}.
\newblock \bibinfo{title}{{Europarl-ST}: A multilingual corpus for speech
  translation of parliamentary debates}, in: \bibinfo{booktitle}{Proc. ICASSP},
  \bibinfo{address}{Barcelona, Spain}.
\bibitem[{Kannan et~al.(2018)Kannan, Wu, Nguyen, Sainath, Chen and
  Prabhavalkar}]{8462682}
\bibinfo{author}{Kannan, A.}, \bibinfo{author}{Wu, Y.},
  \bibinfo{author}{Nguyen, P.}, \bibinfo{author}{Sainath, T.N.},
  \bibinfo{author}{Chen, Z.}, \bibinfo{author}{Prabhavalkar, R.},
  \bibinfo{year}{2018}.
\newblock \bibinfo{title}{An analysis of incorporating an external language
  model into a sequence-to-sequence model}, in: \bibinfo{booktitle}{Proc.
  ICASSP}, \bibinfo{address}{Calgary, AB, Canada}.
\bibitem[{Li et~al.(2019a)Li, Sainath, Pang and Wu}]{Li2019SemisupervisedTF}
\bibinfo{author}{Li, B.}, \bibinfo{author}{Sainath, T.N.},
  \bibinfo{author}{Pang, R.}, \bibinfo{author}{Wu, Z.}, \bibinfo{year}{2019}a.
\newblock \bibinfo{title}{Semi-supervised training for end-to-end models via
  weak distillation}, in: \bibinfo{booktitle}{Proc. ICASSP},
  \bibinfo{address}{Brighton, United Kingdom}.
\bibitem[{Li(2022)}]{SIP-2021-0050}
\bibinfo{author}{Li, J.}, \bibinfo{year}{2022}.
\newblock \bibinfo{title}{Recent advances in end-to-end automatic speech
  recognition}.
\newblock \bibinfo{journal}{APSIPA Transactions on Signal and Information
  Processing} \bibinfo{volume}{11}.
\bibitem[{Li et~al.(2020)Li, Wu, Gaur, Wang, Zhao and Liu}]{li20_interspeech}
\bibinfo{author}{Li, J.}, \bibinfo{author}{Wu, Y.}, \bibinfo{author}{Gaur, Y.},
  \bibinfo{author}{Wang, C.}, \bibinfo{author}{Zhao, R.}, \bibinfo{author}{Liu,
  S.}, \bibinfo{year}{2020}.
\newblock \bibinfo{title}{On the comparison of popular end-to-end models for
  large scale speech recognition}, in: \bibinfo{booktitle}{Proc. Interspeech},
  \bibinfo{address}{Shanghai, China}.
\bibitem[{Li et~al.(2019b)Li, Zhang and Woodland}]{9003837}
\bibinfo{author}{Li, Q.}, \bibinfo{author}{Zhang, C.},
  \bibinfo{author}{Woodland, P.C.}, \bibinfo{year}{2019}b.
\newblock \bibinfo{title}{Integrating source-channel and attention-based
  sequence-to-sequence models for speech recognition}, in:
  \bibinfo{booktitle}{Proc. ASRU}, \bibinfo{address}{Singapore}.
\bibitem[{Li et~al.(2023)Li, Zhang and Woodland}]{LI202312}
\bibinfo{author}{Li, Q.}, \bibinfo{author}{Zhang, C.},
  \bibinfo{author}{Woodland, P.C.}, \bibinfo{year}{2023}.
\newblock \bibinfo{title}{Combining hybrid {DNN-HMM ASR} systems with
  attention-based models using lattice rescoring}.
\newblock \bibinfo{journal}{Speech Communication} \bibinfo{volume}{147},
  \bibinfo{pages}{12--21}.
\bibitem[{Ling et~al.(2022)Ling, Shen, Cai and Ma}]{Ling2021ImprovingPT}
\bibinfo{author}{Ling, S.}, \bibinfo{author}{Shen, C.}, \bibinfo{author}{Cai,
  M.}, \bibinfo{author}{Ma, Z.}, \bibinfo{year}{2022}.
\newblock \bibinfo{title}{Improving pseudo-label training for end-to-end speech
  recognition using gradient mask}, in: \bibinfo{booktitle}{Proc. ICASSP},
  \bibinfo{address}{Singapore}.
\bibitem[{McDermott et~al.(2019)McDermott, Sak and Variani}]{9003790}
\bibinfo{author}{McDermott, E.}, \bibinfo{author}{Sak, H.},
  \bibinfo{author}{Variani, E.}, \bibinfo{year}{2019}.
\newblock \bibinfo{title}{A density ratio approach to language model fusion in
  end-to-end automatic speech recognition}, in: \bibinfo{booktitle}{Proc.
  ASRU}, \bibinfo{address}{Cartagena, Colombia}.
\bibitem[{Meng et~al.(2022a)Meng, Chen, Prabhavalkar, Zhang, Wang, Audhkhasi,
  Emond, Strohman, Ramabhadran, Huang, Variani, Huang and
  Moreno}]{Meng2022ModularHA}
\bibinfo{author}{Meng, Z.}, \bibinfo{author}{Chen, T.},
  \bibinfo{author}{Prabhavalkar, R.}, \bibinfo{author}{Zhang, Y.},
  \bibinfo{author}{Wang, G.}, \bibinfo{author}{Audhkhasi, K.},
  \bibinfo{author}{Emond, J.}, \bibinfo{author}{Strohman, T.},
  \bibinfo{author}{Ramabhadran, B.}, \bibinfo{author}{Huang, W.R.},
  \bibinfo{author}{Variani, E.}, \bibinfo{author}{Huang, Y.},
  \bibinfo{author}{Moreno, P.J.}, \bibinfo{year}{2022}a.
\newblock \bibinfo{title}{Modular hybrid autoregressive transducer}.
\newblock \bibinfo{journal}{ArXiv} \bibinfo{volume}{abs/2210.17049}.
\bibitem[{Meng et~al.(2022b)Meng, Gaur, Kanda, Li, Chen, Wu and
  Gong}]{meng22_interspeech}
\bibinfo{author}{Meng, Z.}, \bibinfo{author}{Gaur, Y.}, \bibinfo{author}{Kanda,
  N.}, \bibinfo{author}{Li, J.}, \bibinfo{author}{Chen, X.},
  \bibinfo{author}{Wu, Y.}, \bibinfo{author}{Gong, Y.}, \bibinfo{year}{2022}b.
\newblock \bibinfo{title}{Internal language model adaptation with text-only
  data for end-to-end speech recognition}, in: \bibinfo{booktitle}{Proc.
  Interspeech}, \bibinfo{address}{Incheon, Korea}.
\bibitem[{Meng et~al.(2021a)Meng, Kanda, Gaur, Parthasarathy, Sun, Lu, Chen, Li
  and Gong}]{9415039}
\bibinfo{author}{Meng, Z.}, \bibinfo{author}{Kanda, N.}, \bibinfo{author}{Gaur,
  Y.}, \bibinfo{author}{Parthasarathy, S.}, \bibinfo{author}{Sun, E.},
  \bibinfo{author}{Lu, L.}, \bibinfo{author}{Chen, X.}, \bibinfo{author}{Li,
  J.}, \bibinfo{author}{Gong, Y.}, \bibinfo{year}{2021}a.
\newblock \bibinfo{title}{Internal language model training for domain-adaptive
  end-to-end speech recognition}, in: \bibinfo{booktitle}{Proc. ICASSP},
  \bibinfo{address}{Toronto, ON, Canada}.
\bibitem[{Meng et~al.(2021b)Meng, Parthasarathy, Sun, Gaur, Kanda, Lu, Chen,
  Zhao, Li and Gong}]{9383515}
\bibinfo{author}{Meng, Z.}, \bibinfo{author}{Parthasarathy, S.},
  \bibinfo{author}{Sun, E.}, \bibinfo{author}{Gaur, Y.},
  \bibinfo{author}{Kanda, N.}, \bibinfo{author}{Lu, L.}, \bibinfo{author}{Chen,
  X.}, \bibinfo{author}{Zhao, R.}, \bibinfo{author}{Li, J.},
  \bibinfo{author}{Gong, Y.}, \bibinfo{year}{2021}b.
\newblock \bibinfo{title}{Internal language model estimation for
  domain-adaptive end-to-end speech recognition}, in: \bibinfo{booktitle}{Proc.
  SLT}, \bibinfo{address}{Shenzhen, China}.
\bibitem[{Ott et~al.(2019)Ott, Edunov, Baevski, Fan, Gross, Ng, Grangier and
  Auli}]{ott2019fairseq}
\bibinfo{author}{Ott, M.}, \bibinfo{author}{Edunov, S.},
  \bibinfo{author}{Baevski, A.}, \bibinfo{author}{Fan, A.},
  \bibinfo{author}{Gross, S.}, \bibinfo{author}{Ng, N.},
  \bibinfo{author}{Grangier, D.}, \bibinfo{author}{Auli, M.},
  \bibinfo{year}{2019}.
\newblock \bibinfo{title}{Fairseq: A fast, extensible toolkit for sequence
  modeling}, in: \bibinfo{booktitle}{Proceedings of NAACL-HLT 2019:
  Demonstrations}, \bibinfo{address}{Minneapolis, MN, USA}.
\bibitem[{Pallet et~al.(1990)Pallet, Fisher and Fiscus}]{115546}
\bibinfo{author}{Pallet, D.}, \bibinfo{author}{Fisher, W.},
  \bibinfo{author}{Fiscus, J.}, \bibinfo{year}{1990}.
\newblock \bibinfo{title}{Tools for the analysis of benchmark speech
  recognition tests}, in: \bibinfo{booktitle}{Proc. ICASSP},
  \bibinfo{address}{Albuquerque, New Mexico, USA}.
\bibitem[{Panayotov et~al.(2015)Panayotov, Chen, Povey and Khudanpur}]{7178964}
\bibinfo{author}{Panayotov, V.}, \bibinfo{author}{Chen, G.},
  \bibinfo{author}{Povey, D.}, \bibinfo{author}{Khudanpur, S.},
  \bibinfo{year}{2015}.
\newblock \bibinfo{title}{{Librispeech: an ASR corpus based on public domain
  audio books}}, in: \bibinfo{booktitle}{Proc. ICASSP}, \bibinfo{address}{South
  Brisbane, Queensland, Australia}.
\bibitem[{Papineni et~al.(2002)Papineni, Roukos, Ward and
  Zhu}]{papineni2002bleu}
\bibinfo{author}{Papineni, K.}, \bibinfo{author}{Roukos, S.},
  \bibinfo{author}{Ward, T.}, \bibinfo{author}{Zhu, W.J.},
  \bibinfo{year}{2002}.
\newblock \bibinfo{title}{{Bleu}: a method for automatic evaluation of machine
  translation}, in: \bibinfo{booktitle}{{ACL}}, \bibinfo{address}{Philadelphia,
  PA, {USA}}.
\bibitem[{Peyser et~al.(2019)Peyser, Zhang, Sainath and
  Wu}]{Peyser2019ImprovingPO}
\bibinfo{author}{Peyser, C.}, \bibinfo{author}{Zhang, H.},
  \bibinfo{author}{Sainath, T.N.}, \bibinfo{author}{Wu, Z.},
  \bibinfo{year}{2019}.
\newblock \bibinfo{title}{Improving performance of end-to-end asr on numeric
  sequences}, in: \bibinfo{booktitle}{Proc. Interspeech},
  \bibinfo{address}{Graz, Austriaa}.
\bibitem[{Post et~al.(2013)Post, Kumar, Lopez, Karakos, Callison-Burch and
  Khudanpur}]{post-etal-2013-improved}
\bibinfo{author}{Post, M.}, \bibinfo{author}{Kumar, G.},
  \bibinfo{author}{Lopez, A.}, \bibinfo{author}{Karakos, D.},
  \bibinfo{author}{Callison-Burch, C.}, \bibinfo{author}{Khudanpur, S.},
  \bibinfo{year}{2013}.
\newblock \bibinfo{title}{Improved speech-to-text translation with the fisher
  and callhome {S}panish-{E}nglish speech translation corpus}, in:
  \bibinfo{booktitle}{Proc. IWSLT}, \bibinfo{address}{Heidelberg, Germany}.
\bibitem[{Rousseau et~al.(2014)Rousseau, Del{\'e}glise and
  Est{\`e}ve}]{rousseau-etal-2014-enhancing}
\bibinfo{author}{Rousseau, A.}, \bibinfo{author}{Del{\'e}glise, P.},
  \bibinfo{author}{Est{\`e}ve, Y.}, \bibinfo{year}{2014}.
\newblock \bibinfo{title}{Enhancing the {TED}-{LIUM} corpus with selected data
  for language modeling and more {TED} talks}, in: \bibinfo{booktitle}{Proc.
  {LREC}}, \bibinfo{address}{Reykjavik, Iceland}.
\bibitem[{Shan et~al.(2019)Shan, Weng, Wang, Su, Luo, Yu and
  Xie}]{Shan2019ComponentFL}
\bibinfo{author}{Shan, C.}, \bibinfo{author}{Weng, C.}, \bibinfo{author}{Wang,
  G.}, \bibinfo{author}{Su, D.}, \bibinfo{author}{Luo, M.},
  \bibinfo{author}{Yu, D.}, \bibinfo{author}{Xie, L.}, \bibinfo{year}{2019}.
\newblock \bibinfo{title}{Component fusion: Learning replaceable language model
  component for end-to-end speech recognition system}, in:
  \bibinfo{booktitle}{Proc. ICASSP}, \bibinfo{address}{Brighton, United
  Kingdom}.
\bibitem[{Shi et~al.(2021)Shi, Yu, Lu, Liang, Feng, Wang, Qian and
  Xie}]{9413386}
\bibinfo{author}{Shi, X.}, \bibinfo{author}{Yu, F.}, \bibinfo{author}{Lu, Y.},
  \bibinfo{author}{Liang, Y.}, \bibinfo{author}{Feng, Q.},
  \bibinfo{author}{Wang, D.}, \bibinfo{author}{Qian, Y.}, \bibinfo{author}{Xie,
  L.}, \bibinfo{year}{2021}.
\newblock \bibinfo{title}{{The accented English speech recognition challenge
  2020: Open datasets, tracks, baselines, results and methods}}, in:
  \bibinfo{booktitle}{Proc. ICASSP}, \bibinfo{address}{Toronto, ON, Canada}.
\bibitem[{Sriram et~al.(2018)Sriram, Jun, Satheesh and
  Coates}]{sriram18_interspeech}
\bibinfo{author}{Sriram, A.}, \bibinfo{author}{Jun, H.},
  \bibinfo{author}{Satheesh, S.}, \bibinfo{author}{Coates, A.},
  \bibinfo{year}{2018}.
\newblock \bibinfo{title}{{Cold Fusion: Training seq2seq models together with
  language models}}, in: \bibinfo{booktitle}{Proc. Interspeech},
  \bibinfo{address}{Hyderabad, India}.
\bibitem[{Tsunoo et~al.(2022)Tsunoo, Kashiwagi, Narisetty and
  Watanabe}]{tsunoo22_interspeech}
\bibinfo{author}{Tsunoo, E.}, \bibinfo{author}{Kashiwagi, Y.},
  \bibinfo{author}{Narisetty, C.P.}, \bibinfo{author}{Watanabe, S.},
  \bibinfo{year}{2022}.
\newblock \bibinfo{title}{{Residual language model for end-to-end speech
  recognition}}, in: \bibinfo{booktitle}{Proc. Interspeech},
  \bibinfo{address}{Incheon, Korea}.
\bibitem[{Variani et~al.(2020)Variani, Rybach, Allauzen and
  Riley}]{Variani2020HybridAT}
\bibinfo{author}{Variani, E.}, \bibinfo{author}{Rybach, D.},
  \bibinfo{author}{Allauzen, C.}, \bibinfo{author}{Riley, M.},
  \bibinfo{year}{2020}.
\newblock \bibinfo{title}{Hybrid autoregressive transducer ({HAT})}, in:
  \bibinfo{booktitle}{Proc. ICASSP}, \bibinfo{address}{Barcelona, Spain}.
\bibitem[{Vaswani et~al.(2017)Vaswani, Shazeer, Parmar, Uszkoreit, Jones,
  Gomez, Kaiser and Polosukhin}]{Vaswani2017}
\bibinfo{author}{Vaswani, A.}, \bibinfo{author}{Shazeer, N.},
  \bibinfo{author}{Parmar, N.}, \bibinfo{author}{Uszkoreit, J.},
  \bibinfo{author}{Jones, L.}, \bibinfo{author}{Gomez, A.N.},
  \bibinfo{author}{Kaiser, L.}, \bibinfo{author}{Polosukhin, I.},
  \bibinfo{year}{2017}.
\newblock \bibinfo{title}{Attention is all you need}, in:
  \bibinfo{booktitle}{Proc. NeurIPS}, \bibinfo{address}{Long Beach, CA, {USA}}.
\bibitem[{Wang et~al.(2019)Wang, Wang and Lv}]{Wang2019AnOO}
\bibinfo{author}{Wang, D.}, \bibinfo{author}{Wang, X.}, \bibinfo{author}{Lv,
  S.}, \bibinfo{year}{2019}.
\newblock \bibinfo{title}{An overview of end-to-end automatic speech
  recognition}.
\newblock \bibinfo{journal}{Symmetry} \bibinfo{volume}{11},
  \bibinfo{pages}{1018}.
\bibitem[{Watanabe et~al.(2018)Watanabe, Hori, Karita, Hayashi, Nishitoba,
  Unno, Soplin, Heymann, Wiesner and Chen}]{Watanabe2018ESPnet}
\bibinfo{author}{Watanabe, S.}, \bibinfo{author}{Hori, T.},
  \bibinfo{author}{Karita, S.}, \bibinfo{author}{Hayashi, T.},
  \bibinfo{author}{Nishitoba, J.}, \bibinfo{author}{Unno, Y.},
  \bibinfo{author}{Soplin, N.E.Y.}, \bibinfo{author}{Heymann, J.},
  \bibinfo{author}{Wiesner, M.}, \bibinfo{author}{Chen, N.},
  \bibinfo{year}{2018}.
\newblock \bibinfo{title}{{ESPnet}: {E}nd-to-end speech processing toolkit},
  in: \bibinfo{booktitle}{Proc. Interspeech}, \bibinfo{address}{Hyderabad,
  India}.
\bibitem[{Watanabe et~al.(2017)Watanabe, Hori, Kim, Hershey and
  Hayashi}]{8068205}
\bibinfo{author}{Watanabe, S.}, \bibinfo{author}{Hori, T.},
  \bibinfo{author}{Kim, S.}, \bibinfo{author}{Hershey, J.R.},
  \bibinfo{author}{Hayashi, T.}, \bibinfo{year}{2017}.
\newblock \bibinfo{title}{Hybrid {CTC}/attention architecture for end-to-end
  speech recognition}.
\newblock \bibinfo{journal}{IEEE Journal of Selected Topics in Signal
  Processing} \bibinfo{volume}{11}, \bibinfo{pages}{1240--1253}.
\bibitem[{Yang et~al.(2022)Yang, Li and Woodland}]{yang2022knowledge}
\bibinfo{author}{Yang, X.}, \bibinfo{author}{Li, Q.},
  \bibinfo{author}{Woodland, P.C.}, \bibinfo{year}{2022}.
\newblock \bibinfo{title}{Knowledge distillation for neural transducers from
  large self-supervised pre-trained models}, in: \bibinfo{booktitle}{Proc.
  ICASSP}, \bibinfo{address}{Singapore}.
\bibitem[{Young et~al.(1994)Young, Odell and Woodland}]{Young1994TreeBasedST}
\bibinfo{author}{Young, S.J.}, \bibinfo{author}{Odell, J.J.},
  \bibinfo{author}{Woodland, P.C.}, \bibinfo{year}{1994}.
\newblock \bibinfo{title}{Tree-based state tying for high accuracy modelling},
  in: \bibinfo{booktitle}{Proc. HLT}, \bibinfo{address}{Plainsboro, USA}.
\bibitem[{Zeineldeen et~al.(2021)Zeineldeen, Glushko, Michel, Zeyer, Schlüter
  and Ney}]{zeineldeen21_interspeech}
\bibinfo{author}{Zeineldeen, M.}, \bibinfo{author}{Glushko, A.},
  \bibinfo{author}{Michel, W.}, \bibinfo{author}{Zeyer, A.},
  \bibinfo{author}{Schlüter, R.}, \bibinfo{author}{Ney, H.},
  \bibinfo{year}{2021}.
\newblock \bibinfo{title}{Investigating methods to improve language model
  integration for attention-based encoder-decoder asr models}, in:
  \bibinfo{booktitle}{Proc. Interspeech}, \bibinfo{address}{Brno, Czechia}.
\bibitem[{Zhang et~al.(2020)Zhang, Lu, Sak, Tripathi, McDermott, Koo and
  Kumar}]{Zhang2020TransformerTA}
\bibinfo{author}{Zhang, Q.}, \bibinfo{author}{Lu, H.}, \bibinfo{author}{Sak,
  H.}, \bibinfo{author}{Tripathi, A.}, \bibinfo{author}{McDermott, E.},
  \bibinfo{author}{Koo, S.}, \bibinfo{author}{Kumar, S.}, \bibinfo{year}{2020}.
\newblock \bibinfo{title}{Transformer transducer: A streamable speech
  recognition model with transformer encoders and {RNN-T} loss}, in:
  \bibinfo{booktitle}{Proc. ICASSP}, \bibinfo{address}{Barcelona, Spain}.
\bibitem[{Zhao et~al.(2022)Zhao, Xue, Parthasarathy, Miljanic and
  Li}]{Zhao2022FastAA}
\bibinfo{author}{Zhao, R.}, \bibinfo{author}{Xue, J.},
  \bibinfo{author}{Parthasarathy, P.}, \bibinfo{author}{Miljanic, V.},
  \bibinfo{author}{Li, J.}, \bibinfo{year}{2022}.
\newblock \bibinfo{title}{Fast and accurate factorized neural transducer for
  text adaption of end-to-end speech recognition models}.
\newblock \bibinfo{journal}{ArXiv} \bibinfo{volume}{abs/2212.01992}.
\bibitem[{Zhao et~al.(2019)Zhao, Li, Wang and Li}]{8682586}
\bibinfo{author}{Zhao, Y.}, \bibinfo{author}{Li, J.}, \bibinfo{author}{Wang,
  X.}, \bibinfo{author}{Li, Y.}, \bibinfo{year}{2019}.
\newblock \bibinfo{title}{{The speechtransformer for large-scale Mandarin
  Chinese speech recognition}}, in: \bibinfo{booktitle}{Proc. ICASSP},
  \bibinfo{address}{Brighton, United Kingdom}.
\bibitem[{Zheng et~al.(2021)Zheng, Liu, Gunceler and
  Willett}]{Zheng2020UsingSA}
\bibinfo{author}{Zheng, X.}, \bibinfo{author}{Liu, Y.},
  \bibinfo{author}{Gunceler, D.}, \bibinfo{author}{Willett, D.},
  \bibinfo{year}{2021}.
\newblock \bibinfo{title}{Using synthetic audio to improve the recognition of
  out-of-vocabulary words in end-to-end {ASR} systems}, in:
  \bibinfo{booktitle}{Proc. ICASSP}, \bibinfo{address}{Toronto, ON, Canada}.
\bibitem[{Zhou et~al.(2022)Zhou, Zheng, Schlüter and Ney}]{9746948}
\bibinfo{author}{Zhou, W.}, \bibinfo{author}{Zheng, Z.},
  \bibinfo{author}{Schlüter, R.}, \bibinfo{author}{Ney, H.},
  \bibinfo{year}{2022}.
\newblock \bibinfo{title}{{On language model integration for RNN transducer
  based speech recognition}}, in: \bibinfo{booktitle}{Proc. ICASSP},
  \bibinfo{address}{Singapore}.

\end{thebibliography}





\end{document}